\documentclass[aps,pra,reprint,amsmath,amssymb,showpacs,groupedaddress,floatfix]{revtex4-1}
\usepackage{mathtools,graphicx,bbold,bm,color,url}
\usepackage[colorlinks,citecolor=blue,urlcolor=blue,linkcolor=blue,hyperindex,breaklinks]{hyperref}

\newcommand{\dd}{\text{d}}

\providecommand{\ket}[1]{\lvert #1 \rangle}

\begin{document}
%\preprint{}

\title{Hawking spectrum for a fiber-optical analog of the event horizon}
\author{David Bermudez}
\email{dbermudez@fis.cinvestav.mx}
\homepage{\\ http://www.fis.cinvestav.mx/{$\sim$}dbermudez/}
%\thanks{I would like to thank Weizmann Institute of Science for their hospitality during the realization of this work.}
\affiliation{Departamento de F\'{\i}sica. Cinvestav, A.P. 14-740, 07000 Ciudad de M\'exico, Mexico}

\author{Ulf Leonhardt}
\email{ulf.leonhardt@weizmann.ac.il}
\homepage{\\ http://ulfleonhardt.weizmann.ac.il}
\affiliation{Department of Physics of Complex Systems. Weizmann Institute of Science, Rehovot 76100, Israel}

%\date{\today}

\begin{abstract}
Hawking radiation has been regarded as a more general phenomenon than in gravitational physics, in particular in laboratory analogs of the event horizon. Here we consider the fiber-optical analog of the event horizon, where intense light pulses in fibers establish horizons for probe light. Then, we calculate the Hawking spectrum in an experimentally realizable system. We found that the Hawking radiation is peaked around group-velocity horizons in which the speed of the pulse matches the group velocity of the probe light. The radiation nearly vanishes at the phase horizon where the speed of the pulse matches the phase velocity of light.
%\noindent Journal reference: \emph{Phys.\ Rev.\ A\ }\textbf{93}, 012115 (2016), DOI: %\href{http://dx.doi.org/10.1103/PhysRevA.93.012115}{10.1103/PhysRevA.93.012115}
\end{abstract}
\pacs{42.50.-p Quantum optics. 42.81.-i Fiber optics. 04.70.Dy Quantum aspects of black holes, evaporation, thermodynamics.}

\maketitle

\section{Introduction}
More than forty years ago Hawking \cite{Haw74,Haw75} predicted that the horizon of a black hole is not black after all, but emits thermal radiation with a characteristic temperature consistent with Bekenstein's black-hole thermodynamics \cite{Bek73}. Since then, Hawking's radiation and Bekenstein's entropy have been the crucial tests for potential quantum theories of gravity, although these tests have remained unproven themselves---there is no experimental evidence for Hawking radiation in astrophysics, and this is likely to remain for the foreseeable future. However, a new approach to Hawking radiation has become experimentally accessible: analogs of gravity. Studying such analogs, we have already gained insights into the trans-Planckian problem that arises due to the infinite frequency shifts at horizons. In analog systems, the frequency shift is limited due to the frequency or wavelength dependence of the wave velocity, i.e., due to dispersion. However, one of the unavoidable consequences of dispersive systems is the loss of strict thermality in the spectrum of Hawking radiation. What exactly is the expected Hawking spectrum for experimentally realizable systems? This is the question we answer here for fiber-optical systems \cite{Leo08}.

Quantum field theory (QFT) tells us that there is a physical state that fills the entire universe; it is the state of absolute darkness, the quantum vacuum. Quantum vacuum is predicted to have physical consequences: the strong gravitational field around a black hole produces Hawking radiation \cite{Haw74,Haw75}; a related phenomenon is the Unruh effect \cite{Unr81}, where an accelerated detector in the Minkowski vacuum measures something more than just vacuum: particles; also, a strong electric field can detach electron-positron pairs from the vacuum, which is known as the Schwinger effect \cite{Sch51}; and finally, the dynamical Casimir effect \cite{Moo70,Casimir11}, where the change in the boundary conditions of an electromagnetic field creates photons.

In this work we will deal exclusively with Hawking radiation, which is considered one of the most secure hypotheses of a future quantum theory of gravity. In fact, it is used to check the viability of new theories \cite{Hel03}. Nevertheless, it rests on two questionable assumptions: first, the derivation of Hawking radiation needs wavelengths shorter than the Planck length, where we expect the known physics to fail, which is known as the trans-Planckian problem; second, assuming that we expect no new physics in those regimes would imply that we cannot use Hawking radiation as a test of new theories. Therefore, the challenge of a theory of quantum gravity is not to reproduce the Hawking radiation hypothesis, but rather to explain what happens to the quantum fields around an event horizon, the boundary that limits a black hole.

Anyone studying Hawking radiation should accept these issues and, if possible, strive to explain them. Yet, Hawking radiation is a good starting point to study the connection between gravity and quantum physics. Its study combines naturally different research areas---gravity, quantum theory, and thermodynamics---but it is still simple enough to be addressed theoretically. There is one problem though: the nearest black hole is thousands of light years away from Earth and, even more, it seems that in the foreseeable future it will not be possible to measure radiation coming from it due to cosmic noise from the cosmic microwave background radiation.

There are several phenomena that surpass our present, and sometimes foreseeable, observational capacities, yet we believe in them due to their strong theoretical support. One way scientists have come up to study these phenomena is via analog systems, where a part of the actual system is replicated with a different one such that its equations are similar. Moreover, analogs have the advantage that they can be designed to be more efficient than the original systems, thus enhancing our capacity to learn from them through the understanding of their similarities and differences.

The first realization that Hawking radiation could be a more general effect was given by Unruh \cite{Unr81}, although it was largely unrecognized at the time. This work included an analog model based on fluid flow. Over time, Physicists have studied the consequences of this work and concluded that Hawking radiation has nothing to do with general relativity \textit{per se}, but that it is a more fundamental phenomenon derived directly from curved-space QFT and is present wherever there is a horizon.

Several analog systems have appeared since then, including water waves \cite{Rou08,Rou10}, Bose-Einstein condensates \cite{Ste14}, chains of superconducting quantum interference devices (SQUIDs) \cite{Par14,LPHH14}, excito-polariton superfluids \cite{SFM11,GC12} and ultra-short laser pulses \cite{Leo08,Fac10}. These systems have something in common: they mimic quantum effects in such a way that it may be possible to measure them in a laboratory. In this work we will focus on quantum-optical analogs using optical fibers \cite{Leo08}.

The first optical-analog proposed the use of slow light \cite{LP00}, although it was later realized that superluminal velocities \footnote{The point where a wave surpasses the speed of light is usually called \textit{phase horizon}, although strictly speaking it is not an horizon. We will avoid this terminology here.} are an essential for particle creation \cite{US12}. Nevertheless, it inspired another optical analog, using light pulses in fibers \cite{Leo08}, which fulfills the conditions for particle creation: superluminal velocities and negative norm in the group-velocity horizon.

After, dispersive systems were studied \cite{Rob11,Rob12,Finazzi13,Finazzi14,Konig15}. These systems keep the creation of particles due to the Hawking effect, but its spectrum would not be thermal anymore, it will rather be strongly dependent on dispersion. In the present work, we employ a recently developed numerical method \cite{RL14} to solve a scattering problem entirely in Fourier space. This method can be applied to the event horizon of the optical analog with a realistic dispersion of the fiber to obtain its scattering spectrum, which we called the Hawking spectrum, which will be strongly non-thermal.

\section{The optical analog of the event horizon} \label{analog}

All the analogs of the event horizon consider the black-hole spacetime as a moving medium, i.e., as a fluid whose movement is caused by gravity. For the optical case the analogy goes one step further: the waves are light waves and the moving fluid is replaced by propagation inside a dielectric material. We will summarize and compare both analogies.

\subsection{Space time as a moving fluid}\label{stasfluid}
As we are interested in the most basic features of black holes, we choose to study the simplest of them: those that only have mass $M$ (no charge $Q$ nor angular momentum $L$). These black holes are described by the Schwarzschild metric $\dd s$, which is the metric of a spherically-symmetric space with a mass $M$ at the origin. This metric is given in Painlev\'e \cite{Pai21}, Gullstrand \cite{Gul22} and Lema\^itre \cite{Lem33} coordinates by:
\begin{equation}
\dd s^2= c^2 \dd t^2-\left( \dd r + \sqrt{\frac{r_S}{r}}c\, \dd t\right)^{2} - r^2\dd\Omega^2,
\label{schw}
\end{equation}
where $\dd \Omega^2 = \dd \theta+\sin^2\theta\, \dd\phi^2$ is the solid-angle element, $(r,\theta,\phi)$ are the spherical coordinates, $c$ is the speed of light in vacuum, and $r_S=2GM/c^2$ is the Schwarzschild radius. As we are interested in the Hawking effect and the fluid analog we will only consider a $1+1$ metric by setting $\dd\Omega=0$.

The light cones (that fulfill $\dd s^2=0$) of this metric have the following trajectories:
\begin{equation}
\frac{\dd t}{\dd r}= \pm \frac{1}{c}\left( 1 \mp \sqrt{\frac{r_S}{r}}\right)^{-1}.
\label{geoeq}
\end{equation}
Consider the behavior of light rays in two different regimes: for $r \gg r_S$, their speed approaches $c$; for $r\rightarrow r_S$ it goes to zero. Then, light rays traveling towards $r_S$ can never reach it (it would take them an infinite time)

\begin{figure}
	\centering
	\includegraphics[width=85mm]{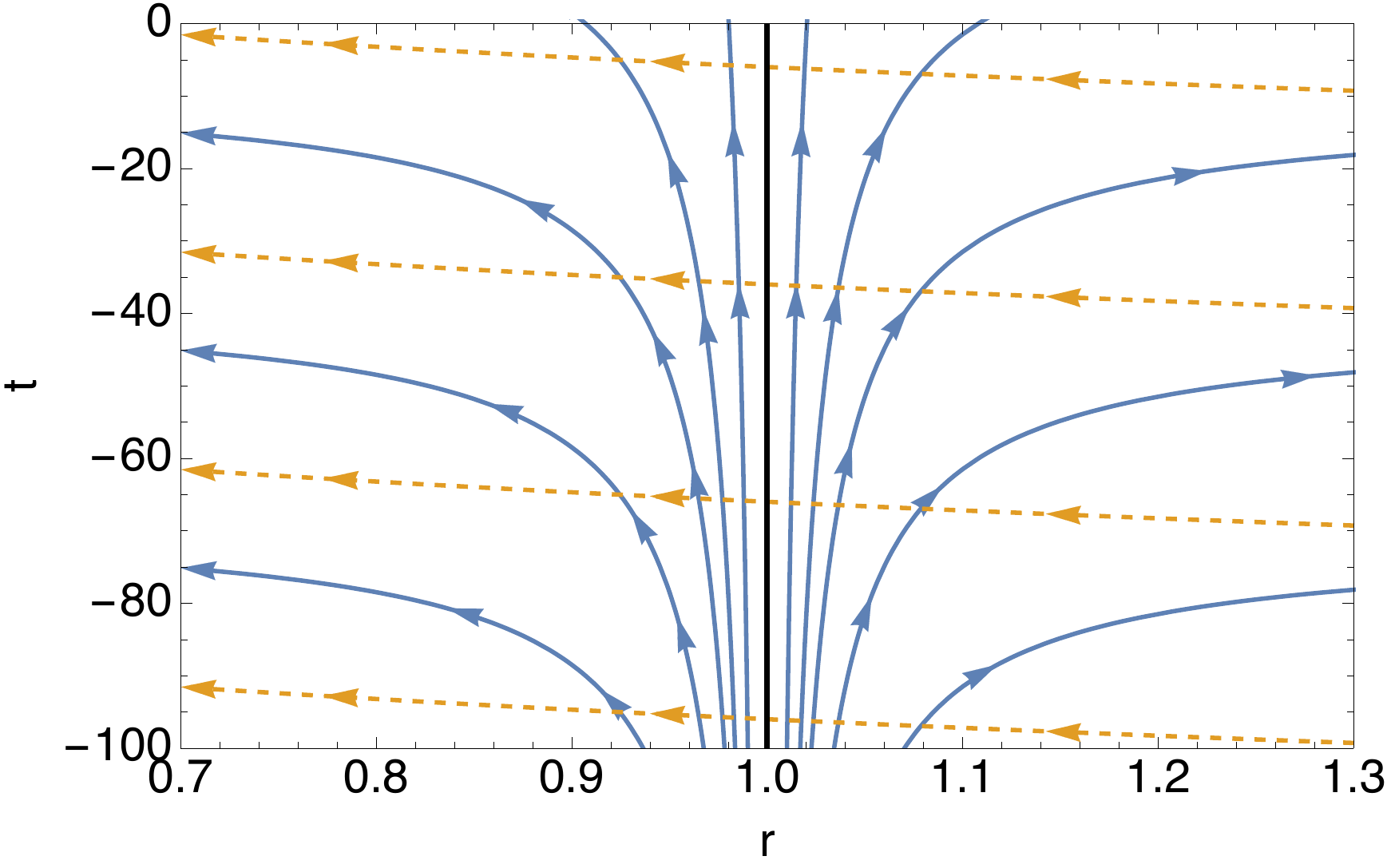}
	\caption{(Color online). Space-time diagram of light-ray trajectories near the horizon given by Eq. \eqref{geoeq}. Straight lines (dashed orange) represent rays traveling with the fluid (copropagating waves), for which there is nothing special in the horizon. Curved lines (solid blue) represent rays traveling against the fluid (counterpropagating waves) and they are split into two at the horizon. The center black-line is the horizon. The radius $r$ is written in terms of $r_S^{-1}$ units.}
	\label{geodesics}
\end{figure}

The plus sign describes light falling in, the minus sign light traveling out, or trying to do so. For $r \gg r_S$ the speed of both rays approaches $c$. For $r \rightarrow r_S$ the speed of the outgoing waves approaches zero, for $r < r_S$ is negative: even outgoing waves are drawn in and fall towards the central singularity. Therefore, the surface $r=r_S$ defines the event horizon \cite{gravity}. The behavior of these two geodesics is shown in Fig. \ref{geodesics}, where the analogy with a moving fluid starts to be useful: light rays moving with the space-time fluid pass the horizon without problem, i.e., there is nothing special for them there. On the other hand, light rays moving against the fluid are not regular at the horizon, as we just saw, their velocity is exactly zero there.

Here we define exactly what we mean by an analog of the event horizon. If we replace the term $-c\sqrt{r_S/r}$ with a general velocity profile $u(r)$ in the (1+1)-dimensional version of Eq. \eqref{schw} we obtain the metric:
\begin{equation}
\dd s^2 = c^2 \dd t^2 - [\dd r - u(r) \dd t]^2,
\label{metric}
\end{equation}
where $c$ is now the speed  of rays with respect to the medium. The point here is not only that the general velocity profile can be different from $\propto r^{-1/2}$ as in the black-hole metric, but also that this profile does not need to be caused by gravity. Hence, this is an analog system in which we generalize both shape and origin of the effects. This is the common proposal in the analog gravity community: some effects caused by gravity are more general and could also have different origins. Specifically, Hawking radiation is a phenomenon from curved-space QFT originated by the event horizon, independently of what causes the horizon (gravity or otherwise). To differentiate this point of view, the new systems are called analogs of the original ones. We should keep in mind this difference (for a review of works in analog gravity see Ref. \cite{LivRev11}).

Let us describe the analogy of Hawking radiation in detail. In 1974, Hawking \cite{Haw74} proposed that the event horizon of a black hole emits thermal radiation consistent with Bekenstein's black hole thermodynamics \cite{Bek73}. Hawking radiation has effective temperature $T$ given by \cite{Haw74,Haw75}:
\begin{eqnarray}
k_B T=\frac{\hbar \kappa}{2\pi},
\label{hre}
\end{eqnarray}
where $k_B$ is Boltzmann's constant and $\kappa$ is the surface gravity. For the astrophysical case $\kappa$ is given by
\begin{equation}
\kappa=\frac{c^3}{4GM},
\label{originalk}
\end{equation}
where $M$ is the mass of the black hole. The same equation \eqref{hre} can be obtained \cite{Leo10,BDL15} for the analog case with the general velocity profile $u(r)$, and
\begin{equation}
\kappa=\left.\frac{\partial u}{\partial r}\right\rvert_\text{horizon}.
\label{approx}
\end{equation}
If we use the velocity profile from the Schwarzschild metric $u(r)=-c\sqrt{r_S/r}$ we obtain Hawking's original formula \eqref{originalk}.

In the astrophysical case, the smaller the mass, the higher the temperature of the Hawking radiation and the stronger its emission \cite{gravity}, but the Chandrasekhar limit is a lower bound for the mass of a star to form a black hole. For an analog system, we can try to increase the change of speed in the velocity profile and thus, improve the production of Hawking radiation \cite{Leo08}.

\begin{figure}
	\centering
	\includegraphics[width=85mm]{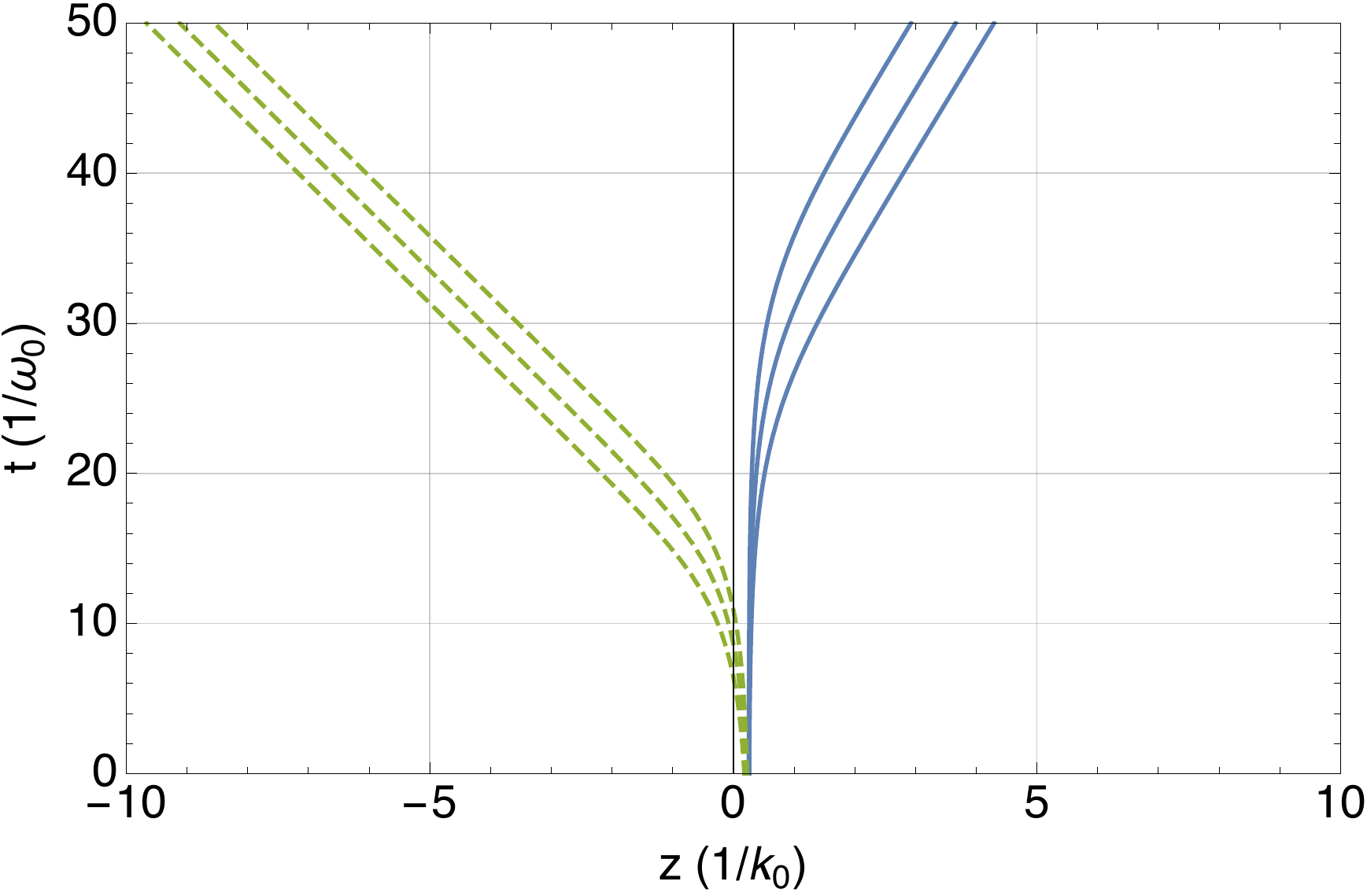}
	\includegraphics[width=85mm]{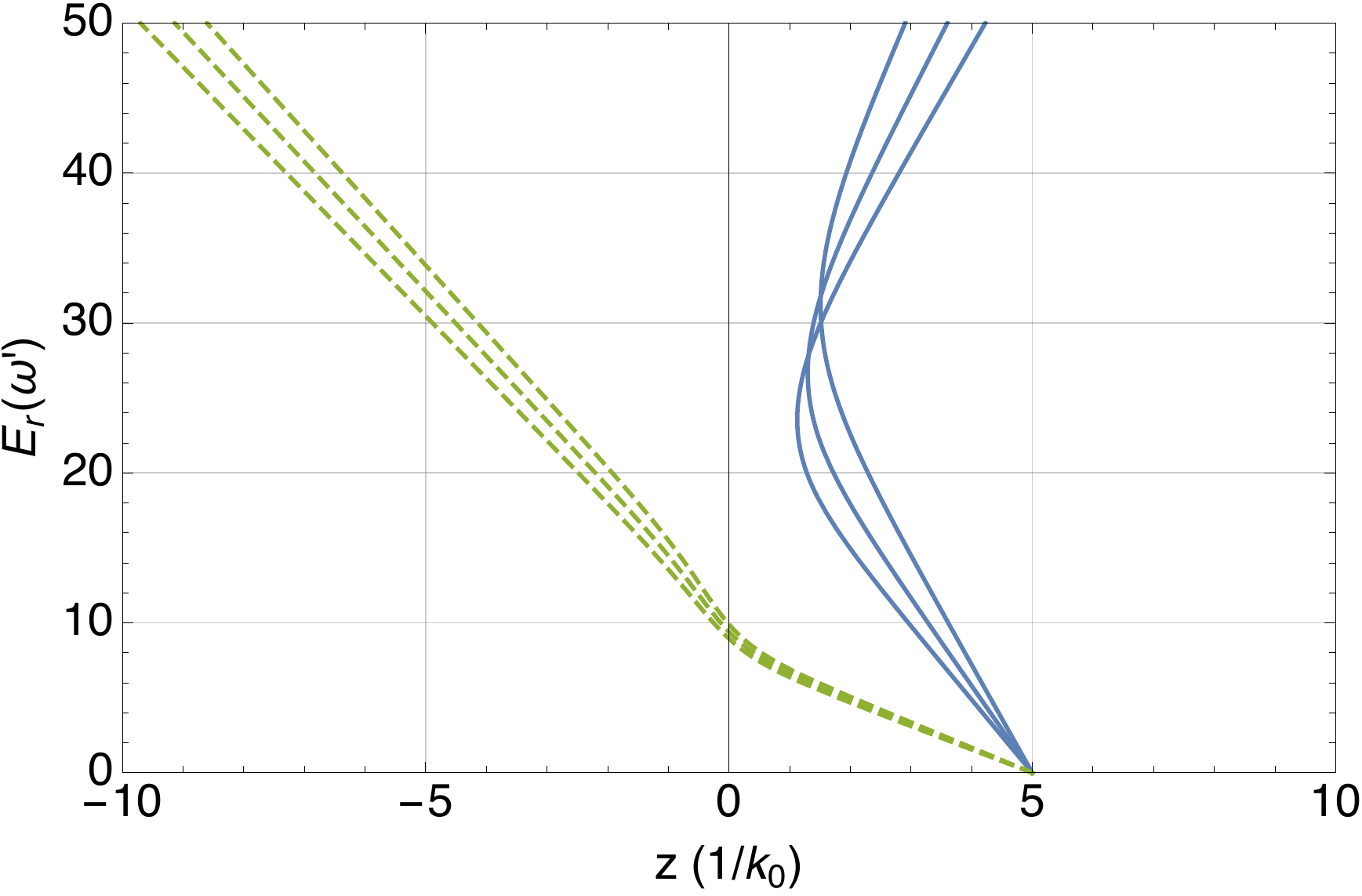}
	\caption{(Color online). Space-time diagram of light rays near the horizon for the velocity profile in Eq. \eqref{velprof} for two cases: dispersionless (up) and dispersion $c(k)$ from Eq. \eqref{dispfluid} (down). In the first case the horizon is well defined, but in the second one is ``fuzzy'' as it depends on $k$. We show three pairs of rays that conserve $\omega-uk$, one with positive (blue) and the other with negative (green) frequency.}
	\label{sts}
\end{figure}

Equation \eqref{metric} is obtained with a dispersionless fluid and the analogy leads to the same equations as the astrophysical case. For example, for a velocity profile $u(z)$ whose geodesics can still be solved analytically is:
\begin{equation}
u(z)=\frac{u_R+u_L}{2}+\frac{u_R-u_L}{2}\tanh\left(\frac{z}{a}\right),
\label{velprof}
\end{equation}
with $a=1/k_0$, $u_L=-1.2\,c_0$, $u_R=-0.8\,c_0$, $k_0$ defines the characteristic length of the dispersion, $u_L$ and $u_R$ are the fluid speed at the far left and right of the horizon, and $c_0$ is the speed of waves in the fluid ($-c_0$ defines the horizons.) The trajectories are illustrated in the top part of Fig. \ref{sts}

However, dispersion is unavoidable in any experimental realization as waves will travel inside real materials---such as silica or a BEC---and when we include it in the theoretical treatment, we lose the exact analogy. For the fluid case, the dispersion is given in terms of $c(k)$, the dependence of its velocity with respect to the wave number $k$. The simplest superluminal dispersion is given by:
\begin{equation}
c(k)=c_0\sqrt{1-\frac{k^2}{k_0^2}}.
\label{dispfluid}
\end{equation}
In this case the geodesics have to be solved numerically and they are shown in the bottom part of Fig. \ref{sts}. By comparing these two plots, we see that in the dispersionless case there is a defined $z$ that separates light rays in two, but in the case with dispersion, the horizon becomes an extended region that depends on the initial $k$ value. We say that the horizon becomes ``fuzzy''.

If we include dispersion, light rays scattered by the horizon will no longer have a thermal spectrum. In this work we investigate the effect of dispersion on the emission and we obtain its Hawking spectrum.

\subsection{Waves in a moving fluid \textit{vs} light in fibers}
We consider light waves propagating in a dielectric material, usually a fiber, in direction $z$ and time $t$ measured in the laboratory frame. Furthermore, these waves can be described by their frequency $\omega$ or their wave number $k$. The dispersion of a fluid is defined by $c(k)$ but for the optical case we are interested in, the dispersion is defined by $n(\omega)$, the dependence of its refractive index on the frequency of light.

Consider a frame moving with velocity $v_0$ following the direction of the waves: the comoving frame. There are two different options to do this. First:
\begin{equation}
z'=z-v_0t,\quad t'=t,
\end{equation}
from which we can obtain
\begin{equation}
\partial_t=\partial_{t'}-v_0\partial_{z'}.
\end{equation}
Thus, the frequency $\omega = i \partial_t$ is not invariant. This leads to a complicated functional form for $n(\omega)$. The second option is
\begin{equation}
\tau=t-\frac{z}{v_0},\quad \zeta=\frac{z}{v_0},
\label{tauzeta}
\end{equation}
which leads to
\begin{equation}
\partial_t=\partial_{\tau}
\end{equation}
and $\omega$ is invariant. For this simplicity, $(\tau,\zeta)$ are the coordinates used by the fiber-optics community \cite{Agrawal}. Let us study how the phase $\phi$ transforms between the laboratory and the comoving frame. In the laboratory frame we have
\begin{equation}
\phi=\int (k \dd z-\omega \dd t),
\end{equation}
and in the comoving frame given by Eq. \eqref{tauzeta} we obtain
\begin{equation}
\phi=-\int (\omega ' \dd \zeta+\omega \dd \tau),
\end{equation}
where we have defined
	\begin{equation}
	\omega'=\omega-v_0 k=\left(1-n(\omega)\frac{v_0}{c}\right)\omega,
	\label{wpw}
	\end{equation}
using the dispersion relation  $k(\omega)=n(\omega)\omega/c$ [in fiber optics \cite{Agrawal} $k(\omega)$ is usually written as $\beta(\omega)$]. Therefore, the role of time is played by $\zeta$,  the propagation distance divided by $v_0$, the retarded time $\tau$ plays the role of distance and, according to the phase, we also have $k$ and $\omega$ played by $-\omega$ and $\omega'$, respectively. In Table \ref{table} we summarize the relation between each quantity in the two frames.

\begin{table}
	\centering
	\begin{tabular}{c | c}
		Fluid & Optics \\
		\hline
		$t$ & $\zeta$ \\
		$z$ & $\tau$ \\ \hline
		$k$ & $-\omega$ \\
		$\omega$ & $\omega'$ \\ \hline
		$c(k)$ & $n(\omega)$
	\end{tabular}
	\caption{Relationship between variables in two analog systems for event horizons: the fluid model and the optical one.}
	\label{table}
\end{table}

In the comoving frame the pulse is at rest and the medium (fiber) travels with speed $-v_0$, effectively creating a moving medium and opening the possibility to establish horizons. Furthermore, due to this change of frame, the speed $v_0$ defines the direction of propagation in the comoving frame: waves traveling slower than $v_0$ appear to be traveling in the direction of the moving medium, as copropagating, while pulses traveling faster than $v_0$ continue to be counterpropagating.

\section{The scattering process} \label{scattering}
One simple way of studying a scattering process is to consider its conservation laws. For a non-relativistic process in a stationary background, the conserved quantities are the frequency and the number of particles. For a relativistic process, the second one is replaced by a more general condition: the conservation of norm. If there are waves with opposite-sign norms, this condition is more like a conservation of charge, e.g., when a neutral particle decays into two particles, one having positive and the other negative charge. The conservation of norm appears in all the pair production processes in particle physics, e.g., when a high-energy photon decays into an electron and a positron, effectively creating two particles, one with positive norm and the other with negative norm.

The quantum vacuum state is defined by the absence of quanta; mathematically, it is the state that fulfils:
\begin{equation}
\hat{a} \ket{0}=0, \quad \forall\, \hat{a},
\end{equation}
where $\hat{a}$ is the annihilation operator and $\ket{0}$ the zero-eigenvalue eigenstate. Both the annihilation operator and the quantum vacuum depend on the choice of the basis for the modes, in a scattering process we can define it with in- and out-modes. Moreover, in QFT when waves change the sign of their norms after scattering, the annihilation operators for the in-modes contain creation operators for the out-modes and vice versa.

The nonequivalence of incoming and outgoing modes is present in all scattering processes; it is not an unusual phenomenon by itself, but usually this results in a conversion from incoming waves to outgoing ones such that the norm of each is conserved. However, if the converted waves have norms with opposite signs, then this process could be an amplification \cite{Unruh11} for both positive- and negative-norm waves. If a positive-norm wave generates a negative-norm component, norm conservation implies that the positive-norm component must grow larger: the wave is amplified.

Furthermore, this process also occurs when the amplitude of the initial wave is small, even if the state is the quantum vacuum. When the vacuum is scattered by a horizon, there is an amplification of this quantum noise and particles can be created \cite{Unruh11}. This is one of the main physical points of Hawking radiation.

The in- and the out-modes form two different sets of orthonormal modes and the transformation between them satisfies the following equation:
\begin{equation}
\phi_{\omega , k_i}^{\text{out},+}= \sum_j \alpha ( \omega ; k_i,k_j ) \phi_{\omega , k_j}^{\text{in},+}+ \sum_j \beta ( \omega ; k_i,k_j ) \phi_{\omega , k_j}^{\text{in},-},
\label{alphabeta}
\end{equation}
where the modes $\phi_{\omega , k_i}$ are all normalized to $\pm 1$ according to their superscripts. Since the out-modes are given by both positive- and negative-norm modes, they combine creation and annihilation operators. On the other hand, amplitudes $\alpha$ and $\beta$ fulfill the following norm-conservation equation:
\begin{equation}
\sum_j | \alpha(\omega ; k_i,k_j) |^2 - \sum_j | \beta(\omega ; k_i,k_j) |^2=1.
\label{normcon}
\end{equation}
Equation \eqref{alphabeta} can be seen as a Bogoliubov transformation, and it is known that its coefficients have interpretation as scattering amplitudes. In general, they have to be calculated numerically. The radiation spectrum of an outgoing mode is the sum of squared amplitudes of the opposite-norm ingoing waves $\beta(\omega;k_i,k_j)$. Therefore, the number of emitted particles $N_{k_i}$ on the $k_i$ branch per unit time per unit frequency is \cite{CJ96}:
\begin{equation}
\frac{\partial^2 N_{k_i} }{\partial \omega\, \partial t}= \frac{1}{2 \pi}\sum_j | \beta(\omega ; k_i,k_j) |^2.
\end{equation}
The radiation coming from this branch of the scattering is the Hawking radiation. Hence, to obtain the Hawking spectrum we need to calculate the scattering amplitudes for the ingoing modes $\vec{\mathcal{A}}^\text{in}$ and for the outgoing ones $\vec{\mathcal{A}}^\text{out}$ for a process that mixes waves of opposite norm, i.e., we must find the scattering matrix $\mathcal{S}$ that fulfills:
\begin{equation}
\vec{\mathcal{A}}^\text{out}=\mathcal{S}\vec{\mathcal{A}}^\text{in},
\label{matrixS}
\end{equation}
considering that the vectors $\vec{\mathcal{A}}^\text{out}$, $\vec{\mathcal{A}}^\text{in}$, and the matrix $\mathcal{S}$ are normalized. The elements of $\mathcal{S}^{-1}$ are the coefficients $\alpha$ and $\beta$ of Eq. \eqref{alphabeta}. The coefficients of $\mathcal{S}$ are the same with the labels ``in'' and ``out'' exchanged.

\section{The calculation method} \label{method}
There are several ways to implement the analogy of the event horizon. From an experimental point of view, the optical analogs are very attractive, because light is a simple quantum object and efficient methods and material are available in quantum optics. Additionally, since its inception quantum optics has offered a reliable testing ground for new theories. In fact, some of the most striking predictions of quantum mechanics have been verified in quantum optics, e.g., entanglement and nonlocality. In this section we describe the method to calculate the Hawking spectrum from a fiber-optical analog of the event horizon \cite{Leo08}.

\subsection{The soliton pulse and its half Fourier transforms}\label{secpulse}
\begin{figure}
	\centering
	\includegraphics[width=85mm]{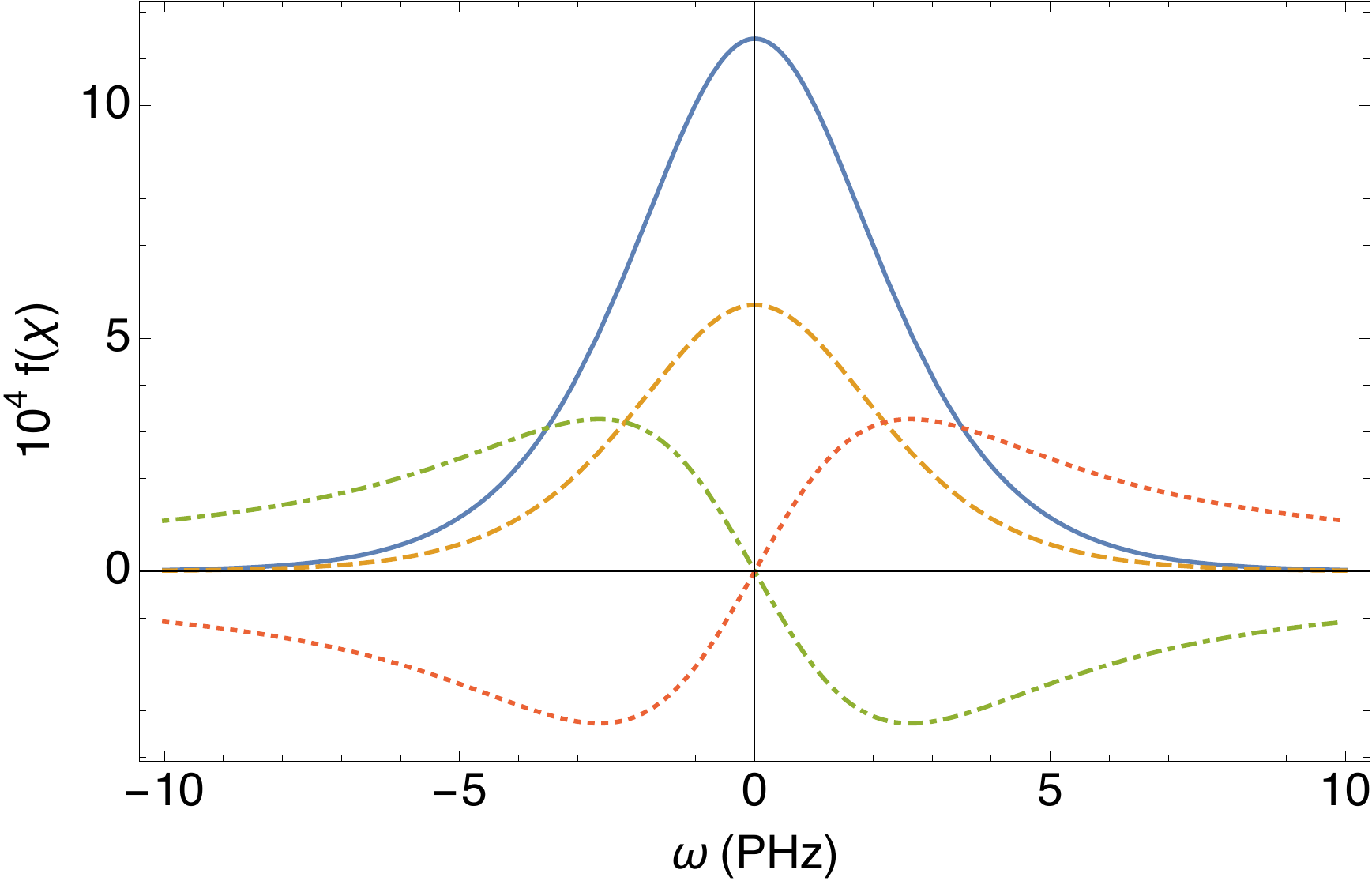}
	\caption{(Color online). The Fourier transform of the sech$(t)$ pulse (solid blue), the real part of the half Fourier transforms (dashed orange), the imaginary parts of left- (dot-dashed green) and right- (dotted red) Fourier transforms.}
	\label{figfts}
\end{figure}

In the optical case, the vacuum state will be scattered by a pulse. It is useful to consider the shortest possible pulses \cite{Leo08}. Currently, there are commercially available short-pulse lasers that produce light in the optical range of $\sim$6\,fs full-width half-maximum (FWHM) duration, i.e., very close to the single-cycle regime. In general these pulses have a bell shape, which we will model as sech, because these pulses have a stable solution that balances the opposite effects of dispersion and nonlinearity when traveling inside a dielectric material, the fundamental soliton. This allows them to travel long distances inside dielectrics without losing their shapes. To form a soliton, the pulse duration and its amplitude cannot be chosen independently, but one is fixed by the other and some fiber parameters. Usually, lasers have a defined duration so we have to tune the intensity to get the fundamental soliton. Their shape is given by:
\begin{equation}
\chi (\tau) = \chi_0\, \text{sech}\left(\frac{\tau}{\tau_0}\right)^2,
\label{soliton}
\end{equation}
where $\chi$ is the nonlinear susceptibility, $\tau_0$ is the pulse duration (usually given in terms of the FWHM time) and $\chi_0$ is fixed by $\tau_0$ and some fiber parameters \cite{Agrawal}. In the Appendix we present the full integral method to calculate the scattering matrix. To apply this method we need the Fourier transform and two half Fourier transforms (left and right) for the pulse. For Eq. \eqref{soliton}, the Fourier transform is given by:
\begin{equation}
\widetilde{\chi}(\omega)=\chi_0\, \pi \tau_0^2 \omega\, \text{csch}\left(\frac{\pi \tau_0 \omega}{2}\right),
\label{FT}
\end{equation}
where the limit $\omega\rightarrow 0$ should be calculated with care and it is equal to $2 \chi_0  \tau_0$. The two half Fourier transforms are
\begin{subequations}
	\begin{align}
	\widetilde{\chi}^L(\omega)&=\chi_0 \tau_0 \omega \left[1-i\frac{\tau_0 \omega}{2}(H_{i\tau_0\omega/4}-H_{-1/2+i \tau_0\omega/4})\right],\\
	\widetilde{\chi}^R(\omega)&=\chi_0 \tau_0 \omega \left[1+i\frac{\tau_0 \omega}{2}(H_{-i\tau_0\omega/4}-H_{-1/2-i \tau_0\omega/4})\right],
	\end{align}
	\label{halfFT}
\end{subequations}
\hspace{-1mm}where $H_n$ is the harmonic number function or, more exactly, its generalization for continuous complex values, which is usually defined through the $\digamma$ function $\Psi_0$ and the Euler-Mascheroni constant $\gamma$ as (see Ref. \cite{CG96} for more details):
\begin{equation}
H_n=\gamma+\Psi_0(n+1).
\end{equation}
In Fig. \ref{figfts} we show the Fourier transform and the two half Fourier transforms given by Eqs. \eqref{FT} and \eqref{halfFT}.

\subsection{Modeling the dispersion relation}
When electromagnetic waves travel inside a medium, they interact with the bound electrons such that their velocity depends on the optical frequency $\omega$ of the wave. This is modeled by the material dispersion and, far from resonances, it is usually well approximated by any of the employed models: Lorentzian, Drude, Debye, or Sellmeier (the most commonly used). For a list of Sellmeier coefficients of different materials see Ref. \cite{refindex}. In our case, we consider optical fibers because their geometry modifies further the material dispersion, giving us also some geometrical parameters to be modeled for convenience.

In the study of optical fibers one usually works with a dispersion relation of a related function $\beta$, which is the wave number $k$ when it depends on $\omega$. It is defined by:
\begin{equation}
\beta(\omega)=n(\omega)\frac{\omega}{c}=\sum_{j=0}^{\infty}\frac{\beta_j}{j!} (\omega-\omega_0)^j,
\label{betat}
\end{equation}
where the last equality is just the Taylor expansion of $\beta$ around a given frequency $\omega_0$. This is a very common way of studying its effect and it gives some mathematical advantages. Also, the physical meaning of the first terms of the expansion is known: $\beta_1$ is the inverse of the group velocity, $\beta_2$ is the group-velocity dispersion (GVD), and $\beta_3$ is the third-order dispersion (TOD) \cite{Agrawal}.

\begin{figure}
	\centering
	\includegraphics[width=85mm]{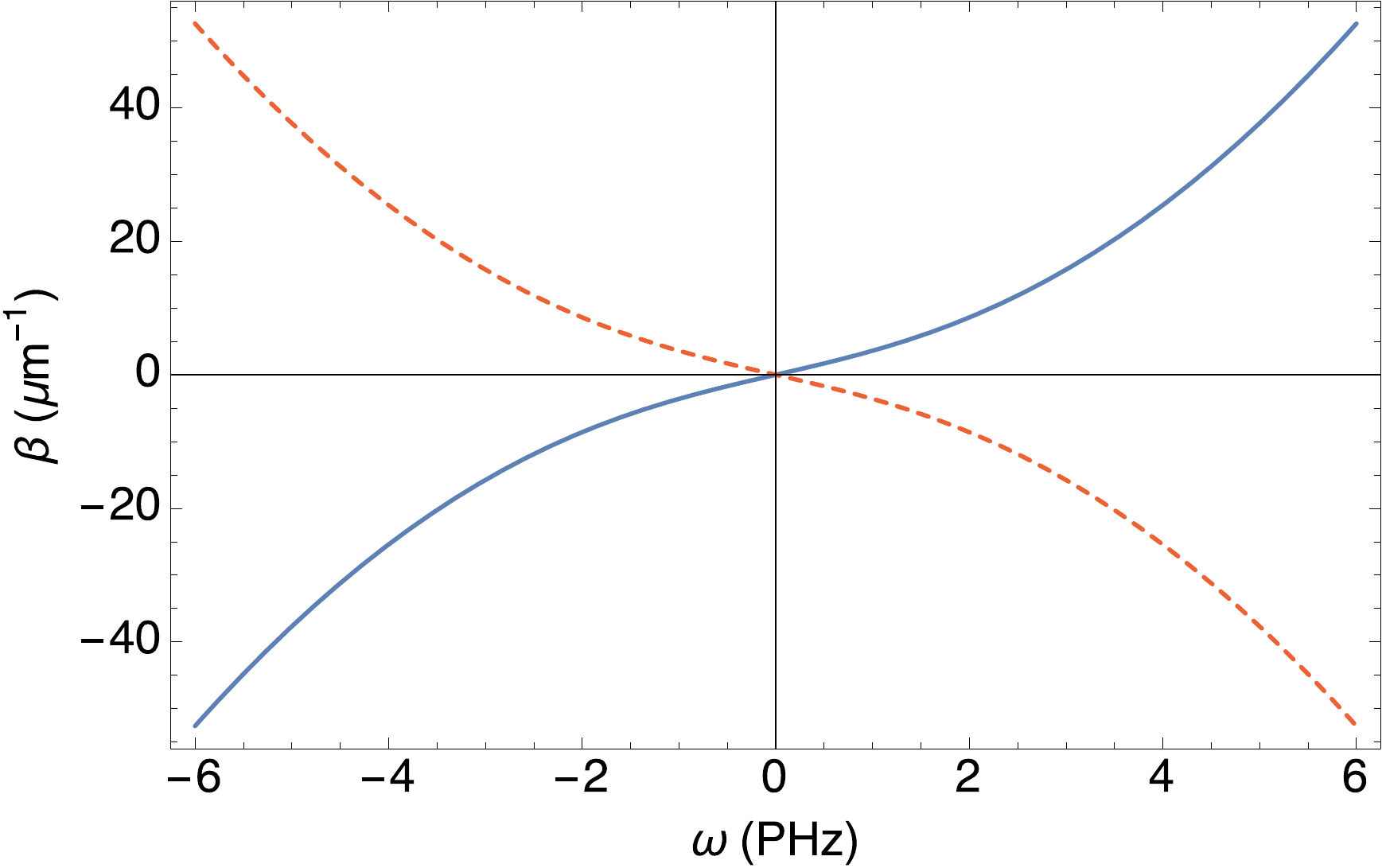}
	\caption{(Color online). Dispersion relation $\beta(\omega)$ for counterpropagating waves using the model in Eq. \eqref{beta} with two terms (solid blue). We also show the negative of this function (dashed red) that is useful to match the negative frequencies.}
	\label{figbeta}
\end{figure}

In this work we are looking for a model that includes the essential properties for modeling the dispersion of light in a fiber and that creates the analog of the event horizon in the optical regime. To do that, we can approximate $\beta^2(\omega)$ by the following equation
\begin{equation}
\beta^2(\omega)=\frac{\omega^2}{c^2}(b_1+b_2\omega^2),
\label{beta}
\end{equation}
where $b_1$ and $b_2$ are parameters. The dispersionless case is obtained with $b_1=1$ and $b_2=0$. Also, the sign of $b_2$ determines if we have a subluminal or superluminal dispersion. This is a simple approximation, but it contains enough detail for modeling the main part of the optical-fiber dispersion, which will be an inflexion point. In Fig. \ref{figbeta} there is a plot of the usual shape of a subluminal dispersion $\beta(\omega)$ for optical fibers from Eq. \eqref{beta}.

\begin{figure}
	\centering
	\includegraphics[width=85mm]{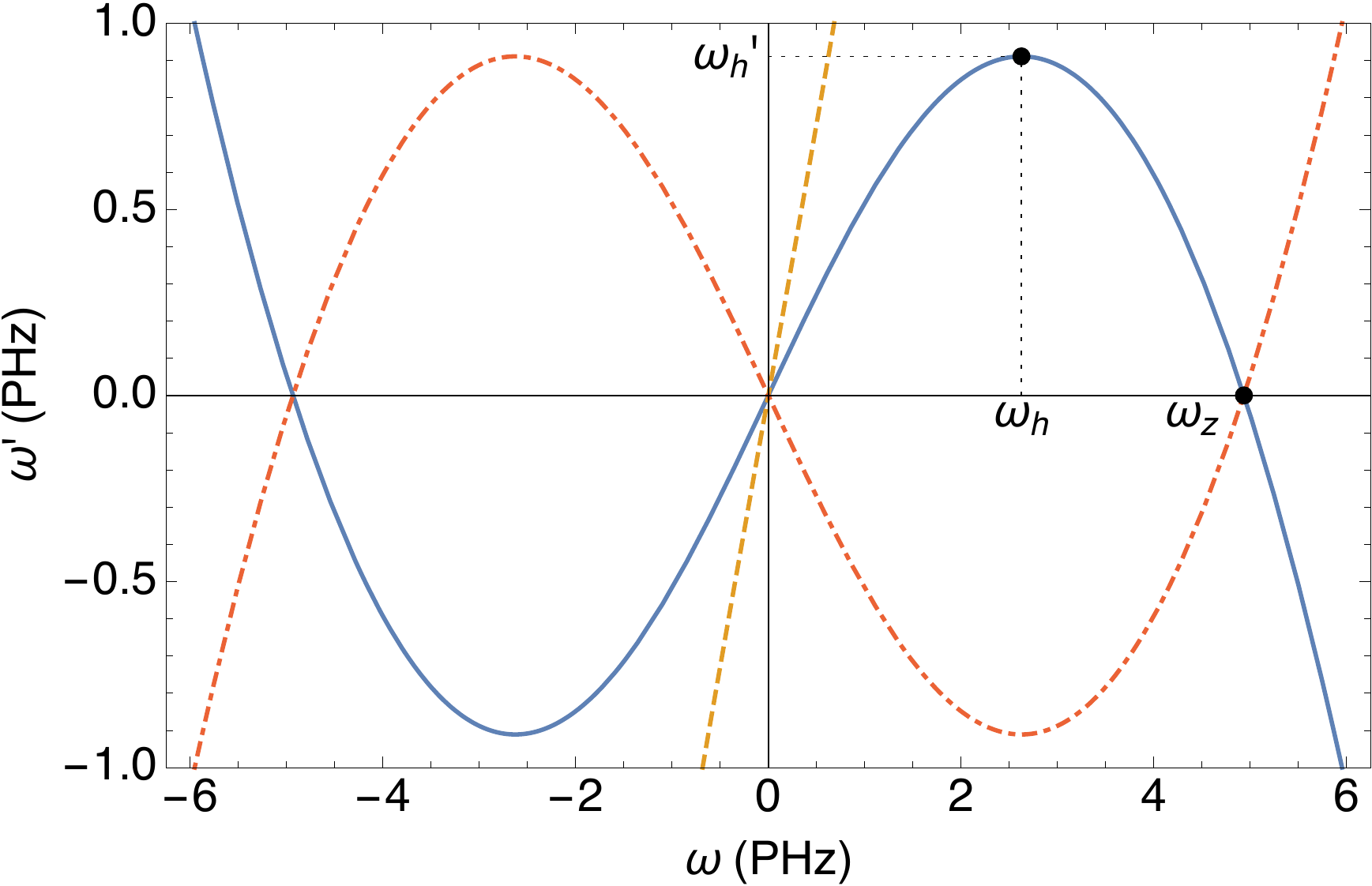}
	\caption{(Color online). Dispersion relation in the comoving frame $\omega'=\omega'(\omega)$ from Eq. \eqref{wpw} and $\beta(\omega)$ from Eq. \eqref{beta}. We show the function for the copropagating waves (solid blue), its negative (dot-dashed red) to match with negative-frequency waves, and the counterpropagating waves (dashed orange). We also show the two points of interest: the phase horizon ($\omega_z,0$) and the group horizon ($\omega_h,\omega_h'$).}
	\label{figwprime2}
\end{figure}

We would like to express parameters $b_1$ and $b_2$ in terms of others more physically meaningful. For this, we have two special points to choose from (see Fig. \ref{figwprime2}.) One is the phase horizon or zero-frequency point, where $\omega'=0$ \footnote{We must remark that $\omega'$ refers to the frequency in the comoving frame and does not represent a derivative.}; the other fulfills $ \dd \omega' /\dd \omega=0$ and we will call it group-velocity horizon or simply the horizon for reasons that will become clear soon. We describe the points in the dispersion by their frequency in the laboratory frame $\omega$ as it is single-valued, while the frequency in the comoving frame is not.

There are two unknown parameters ($b_1$ and $b_2$), and we can obtain a system of two equations by the two conditions for the values of $\omega'=0$ and $ \dd \omega' /\dd \omega=0$ at the points of interest. For the phase horizon $\omega_z$ we have
\begin{equation}
\omega'(\omega_z)=0,\quad 	\left.\frac{\dd \omega'}{\dd \omega}\right|_{w=w_z} =-\epsilon,
\end{equation}
where $\epsilon$ is the new parameter that now has physical meaning: it is the negative derivative of $\omega'$ evaluated at the phase horizon. The solution of the system is:
\begin{equation}
b_1=\frac{c^2}{u^2}(1-\epsilon),\quad b_2=\frac{c^2}{u^2}\frac{\epsilon}{w_z^2}.
\end{equation}
For the horizon $\omega_h$, the system is given by
\begin{equation}
\omega'(\omega_h)=\omega_h',\quad 	\left.\frac{\dd \omega'}{\dd \omega}\right|_{w=w_h} =0,
\end{equation}
where $\omega_h'$ is the new physical parameter, which is the frequency of the horizon in the comoving frame (the other is $\omega_h$ itself). The solution of the new system is:
\begin{subequations}
\begin{align}
b_1&=\frac{c^2}{u^2}\left(1-\frac{\omega_h'}{\omega_h}\right)\left(1-2\frac{\omega_h'}{\omega_h}\right), \\
b_2&=\frac{c^2}{u^2}\frac{\omega_h'}{\omega_h^3}\left(1-\frac{\omega_h'}{\omega_h}\right),
\end{align}
\end{subequations}
so the new parametrization is given in terms of $\omega_h$ and $\omega_h'$. This parametrization is not only physical, but also much closer to the analog event horizon in optical fibers.

In Fig. \ref{figwprime2} we show the same dispersion as in Fig. \ref{figbeta} but now in the comoving frame. The two points $\omega_h$ and $\omega_h'$ are also shown, as well as the negative of the dispersion function, as it is useful to obtain the matching conditions for negative frequencies by only looking at positive ones \cite{Neg12}. We choose the parameters $w_h=2.62645 \text{ PHz}$ and $w_h'= 0.91108 \text{ PHz}$, which gives not only the right order of magnitude but also a good agreement with possible experimental materials.

\subsection{Kerr effect inside an optical fiber}
We need a property that changes the speed of the waves in the medium to obtain the analog. To do this we use a fairly common effect in nonlinear optics \cite{Boyd,Agrawal}: the Kerr effect, which is a nonlinear phenomenon and, as such, needs relatively high pulse-intensities. The change of the refractive index of a fiber due to Kerr effect is
\begin{equation}
n_\text{eff}^2(\omega,t)=n^2(\omega) +\chi(\omega,t),
\label{n2}
\end{equation}
where $n$ is the refractive index of the fiber at rest and $\chi$ is the nonlinear susceptibility, which is proportional to the pulse intensity with the constants of proportionality given by the material response. This is why in Eq. \eqref{soliton} we already wrote the pulse in terms of $\chi$. Initially, $n=n(\omega)$ depends on the frequency but it is constant along the fiber. We obtain $n(\omega)$ from $\beta(\omega)$ through Eq. \eqref{betat} with the model we described in the previous section. Due to the Kerr effect, the refractive index also depends on time as in Eq. \eqref{n2}, because pulses traveling along the fiber change the dispersion. This effect is usually negligible, but in the case where the phase or group velocity of the waves are very close to the ones of the pulse they may change significantly. If we approximate to first order, Eq. \eqref{n2} becomes
\begin{equation}
n_\text{eff}(\omega,t) \simeq n(\omega) +\frac{\chi(\omega,t)}{2n(\omega)}= n(\omega) +\delta n(\omega,t).
\label{n2a}
\end{equation}

\subsection{Dispersion relation in the comoving frame}
The final step to achieve the analogy with the moving fluid is to write down these equations in the comoving frame; the differential equation that appears in the propagation, the nonlinear Schr\"odinger equation (NLSE) \cite{Agrawal,Shalva13,Shalva14,Shalva14b}, is usually solved this way. We consider the Doppler effect in the change of frame. Thus, the frequency $\omega'$ in the comoving frame is
\begin{align}
\omega' (\omega) &= \omega\mp u\beta_\text{eff}(\omega) = \omega\mp \omega \frac{u}{c} n_\text{eff}(\omega)
=\omega\left(1 \mp\frac{n_\text{eff}(\omega)}{n_g(\omega_0)} \right)\nonumber\\
&\simeq \omega\left(1\mp \frac{n(\omega)+\delta n(\omega)}{n_g(\omega_0)} \right),
\label{wfun}
\end{align}
where we drop the explicit dependence on $t$ in Eq. \eqref{n2a} as it only appears as a parameter, $\omega$ is the laboratory frequency and $n_g(\omega_0)=c/u$ is the group velocity of the pulse. The sign of the dispersion corresponds to waves traveling with the pulse (copropagating, negative sign) or against it (counterpropagating, positive sign). In the comoving frame, the phase velocity is $\omega'/k$ and the group velocity is given by $ \dd \omega' /\dd k$. The shape of the dispersion for counter- and copropagating waves can be seen in Fig. \ref{figwprime2}.

We will see that if the soliton is strong enough, a probe pulse that interacts with it can surpass the group velocity of the fiber for certain frequencies and create an event horizon for those frequencies.

\subsection{Analogue of the event horizon}\label{study}
In this section we study more closely the $\omega_h$ frequency, which fulfills $v_g(\omega_h)=0$, establishing what is known as a \textit{group-velocity horizon}.

\begin{figure}
	\centering
	\includegraphics[width=85mm]{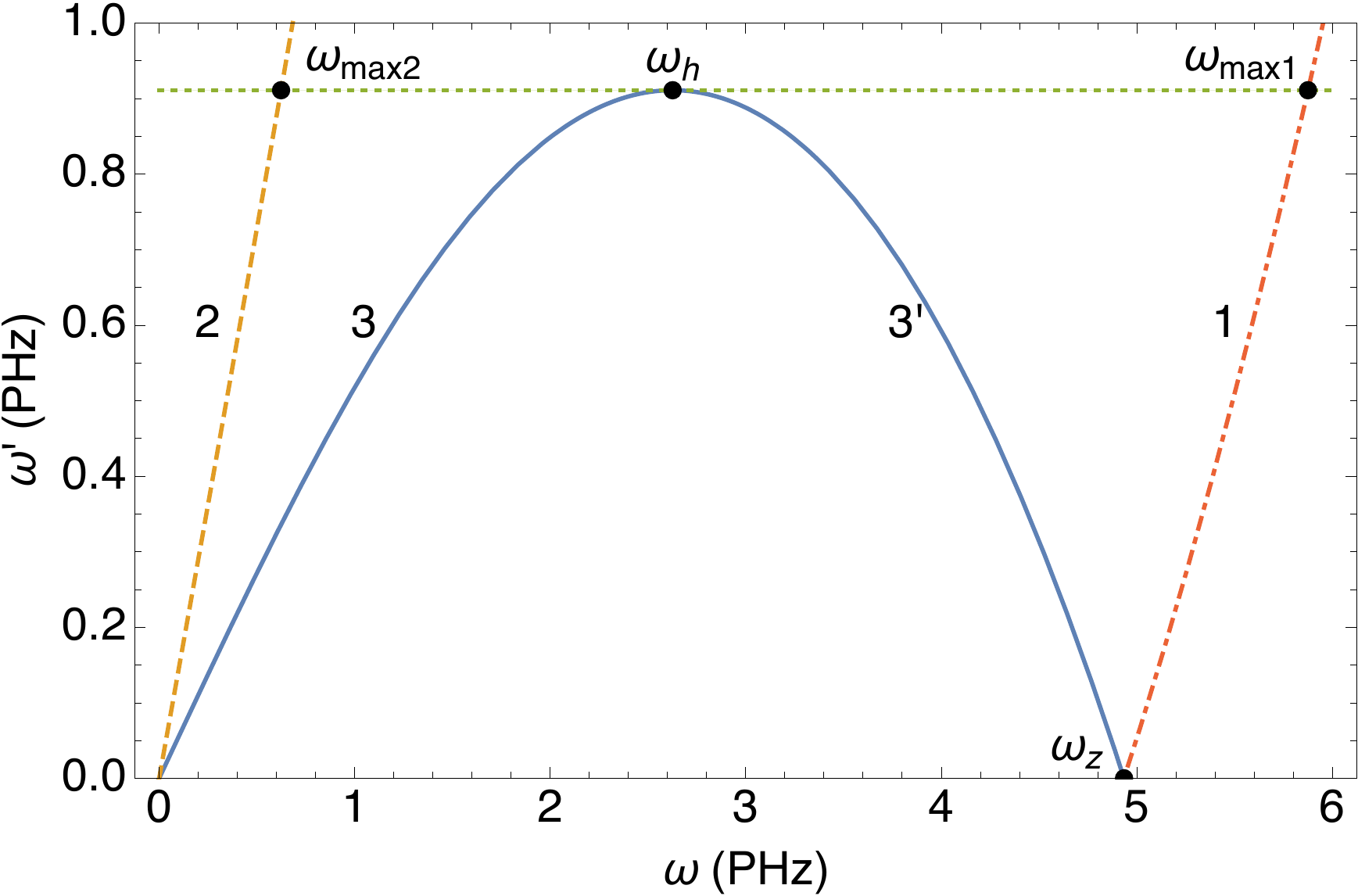}
	\caption{(Color online). The dispersion relation for counterpropagating (solid blue) and copropagating waves (dashed orange); the negative of the counterpropagating (dot-dashed red) dispersion is also plotted as it indicates the matching with the negative frequency. The horizontal line (dotted green) represents the conservation of $\omega'$ and the points shown satisfy the matching conditions for the unperturbed system. The labels 1, 2, 3, and $3'$ correspond to the identification of the modes for the numerical solution.}
	\label{figwpwfull}
\end{figure}

A scattering process combines different frequencies in the laboratory frame, but in the comoving frame $\omega'$ is conserved. So, all the scattered waves have the frequency $\omega'$ in the comoving frame. For our model of dispersion, we will have three modes in $\omega$ (and four different values) where the waves could be scattered to \cite{Parentani11,Parentani11b}, which are shown in Fig. \ref{figwpwfull}. Here we define counter- and copropagating according to the direction in the comoving frame, which opposes the direction of the pulse in the laboratory frame. This is to be in accord with the fluid model.
\begin{itemize}
	\item Mode 1 describes the negative-frequency waves, which we obtained by naturally extending the dispersion relation to negative values, taking advantage of the fact that $\beta(\omega)$ is an odd function. 
	\item Mode 2 contains the copropagating waves, with the dispersion given by Eq. \eqref{wfun} with a negative sign.
	\item Mode 3 describes the counterpropagating waves, with a positive sign in Eq. \eqref{wfun}. It must be split in two due to the existence of the horizon, 3 and $3'$, in order to conserve the uniqueness in the matrix $\mathcal{S}$ from Eq. \eqref{matrixS}.
\end{itemize}

Figure \ref{figwpwfull} illustrates the modes. There we draw a horizontal line in $\omega'=\omega'_h$, which marks the conservation of $\omega'$ and allows us to see that a scattering process from $\omega'_h$ could lead from $\omega_h$ to two other accessible frequencies in the laboratory frame: $\omega_{\text{max}1}$, which is the negative-frequency matching and $\omega_{\text{max}2}$, which is the copropagating one. Usually, the amount of scattering to this last mode is very small because it travels in the opposite direction of the initial waves, but it is included in the calculation for completeness. From the figure we can also see that any other value of $\omega'<\omega_h'$ leads to two possible values for mode 3 (which we called 3 and $3'$). The transition from one to the other allows the input modes to be converted into outgoing modes and we will see that the consequence is the creation of particles (for the black-hole and white-hole horizons). Therefore, an essential ingredient for Hawking radiation is the group-velocity horizon.

\begin{figure}
	\centering
	\includegraphics[width=85mm]{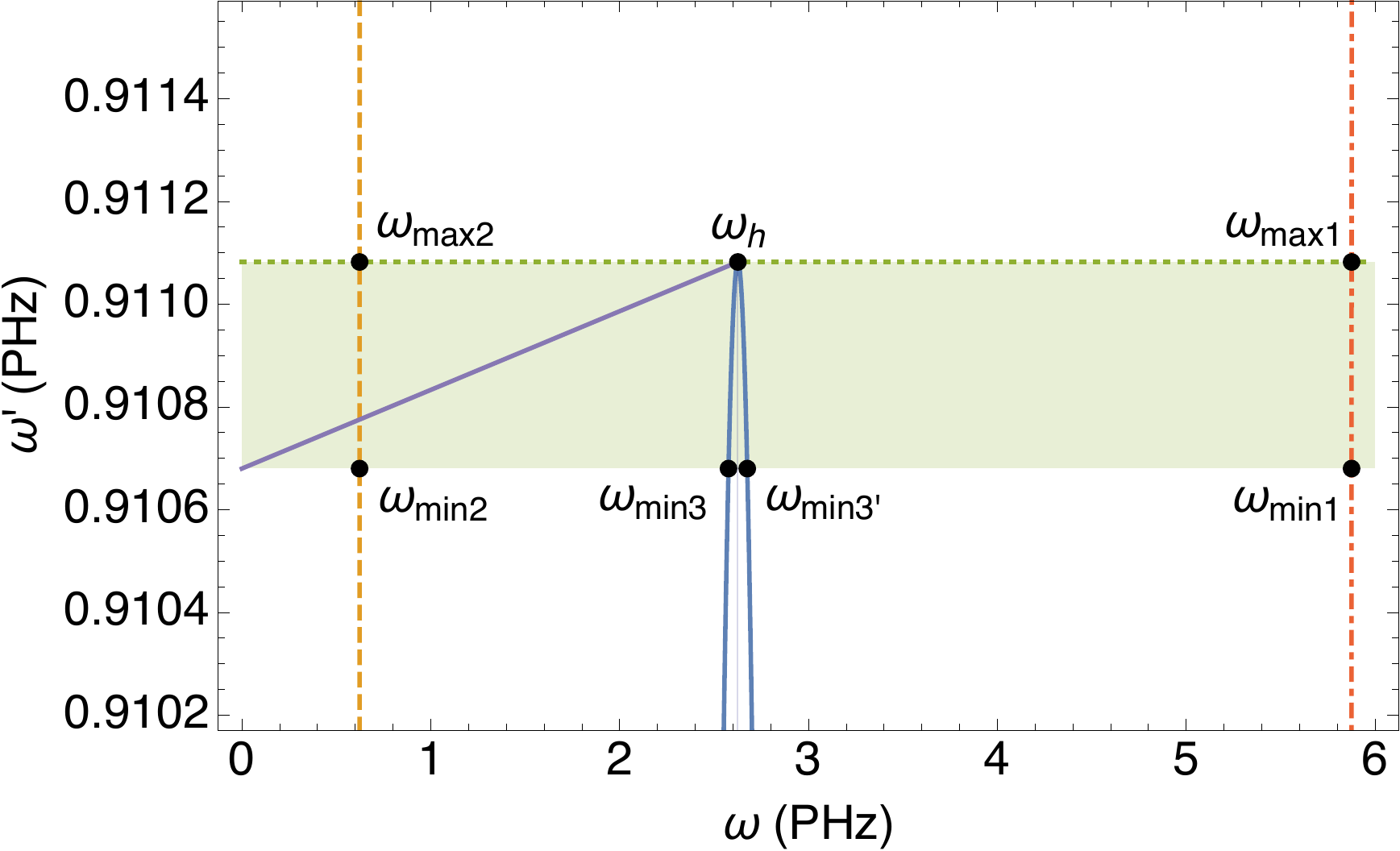}
	\caption{(Color online). Close up on $\omega'_h$ from Fig. \ref{figwpwfull}. The diagonal purple line shows the maximum slope reached by the pulse with $\delta n_\text{max}$ and defines the edge of frequencies $\omega'_\text{min}$ from the three modes that are able to reach the horizon due to the Kerr effect when we consider the soliton. This range is shown in green shadow.}
	\label{figwpwclose}
\end{figure}

On the other hand, as the pulse travels through the fiber, it creates an effective moving medium due to the change of refractive index given by $\delta n(\omega)$. Then, while the pulse passes through a certain fixed point in the comoving frame [according to Eq. \eqref{tauzeta}, this point is moving with speed $v_0$ with respect to the laboratory frame], $\delta n(\omega)$ varies from zero to $\delta n_\text{max}$, defined by the peak intensity of the pulse, i.e., where the susceptibility reaches its maximum, $\chi (\omega,t) =\chi_0$; and then from $\delta n_\text{max}$ to zero when the pulses completes its passing. Due to the Kerr effect, the pulse is able to ``push'' frequencies in the comoving frame that range from $\omega'_h$ to $\omega'_\text{min}$ to the horizon. The frequencies that are closer to the horizon only need a little help from the pulse, while the limit frequency $\omega'_\text{min}$ is set by the peak intensity of the pulse $\delta n_\text{max}$. From Eq. \eqref{wfun} we have:
\begin{equation}
\omega_\text{min}'=\omega_h'-\omega_h\delta n_\text{max},
\end{equation}
which is represented by the diagonal line in Fig. \ref{figwpwclose}.

The other modes also have a new range of frequencies that are able to reach the horizon, not only the points $\omega_\text{max1}$ and $\omega_\text{max2}$. These frequencies reach the horizon in the comoving frame but their laboratory frequencies are still very different and these are the ones that would be measured with a detector in a laboratory. We see in Fig. \ref{figwpwclose} that even though in the comoving frame $\omega'_\text{min}$ is very close to $\omega'_h$, they have very different frequencies in the laboratory frame $\omega$ (see the scales for $\omega'$ and $\omega$ in Fig. \ref{figwpwclose}), thus facilitating their detection with the usual tools of optics laboratories. For our model, the frequencies that can get to the horizon are all in the optical range of the spectrum where there are commercially available detectors, which makes this experiment feasible. In this case, the frequencies are inside [$\omega_\text{min2}$, $\omega_{\text{max2}}$], [$\omega_{\text{min3}}$, $\omega_{\text{min3'}}$] and [$\omega_{\text{min1}}$, $\omega_{\text{max1}}$], as marked in the Fig. \ref{figwpwclose}.

We must remark that we are considering a pulse, which means that $\delta n$ changes twice: $0\rightarrow \delta n_\text{max}$ and  $\delta n_\text{max}\rightarrow 0$, both can create an horizon: in front of the pulse it is a black-hole horizon and in the back a white-hole horizon (the time-reversed equivalent of a black hole). Our method considers the radiation spectrum from the black-hole--white-hole pair that is the most likely configuration of the optical analogs to be measured in the laboratory.

\section{Hawking spectrum}\label{hawking}
In this section we obtain the Hawking spectrum for a fiber using the numerical method described in the Appendix. Then we analyze the numerical errors of the method by checking the norm-conservation.

The algorithm gives the number of photons per unit $t$ per unit $\omega'$ per pulse. The experimentally available quantity is actually the number of photons per unit of time around a certain $\omega$. In order to do that we need to consider the interaction distance, the change between $\omega'$ and $\omega$ and the repetition rate of the laser. Basically, this last one is the most important figure, as all the others will keep the order of magnitude of the result for typical values of an experiment. Therefore, to simplify matters we just multiply our result by the repetition rate, which we considered to be 80 MHz.

\subsection{Numerical results}\label{numerical}
\begin{figure}
	\centering
	\includegraphics[width=85mm]{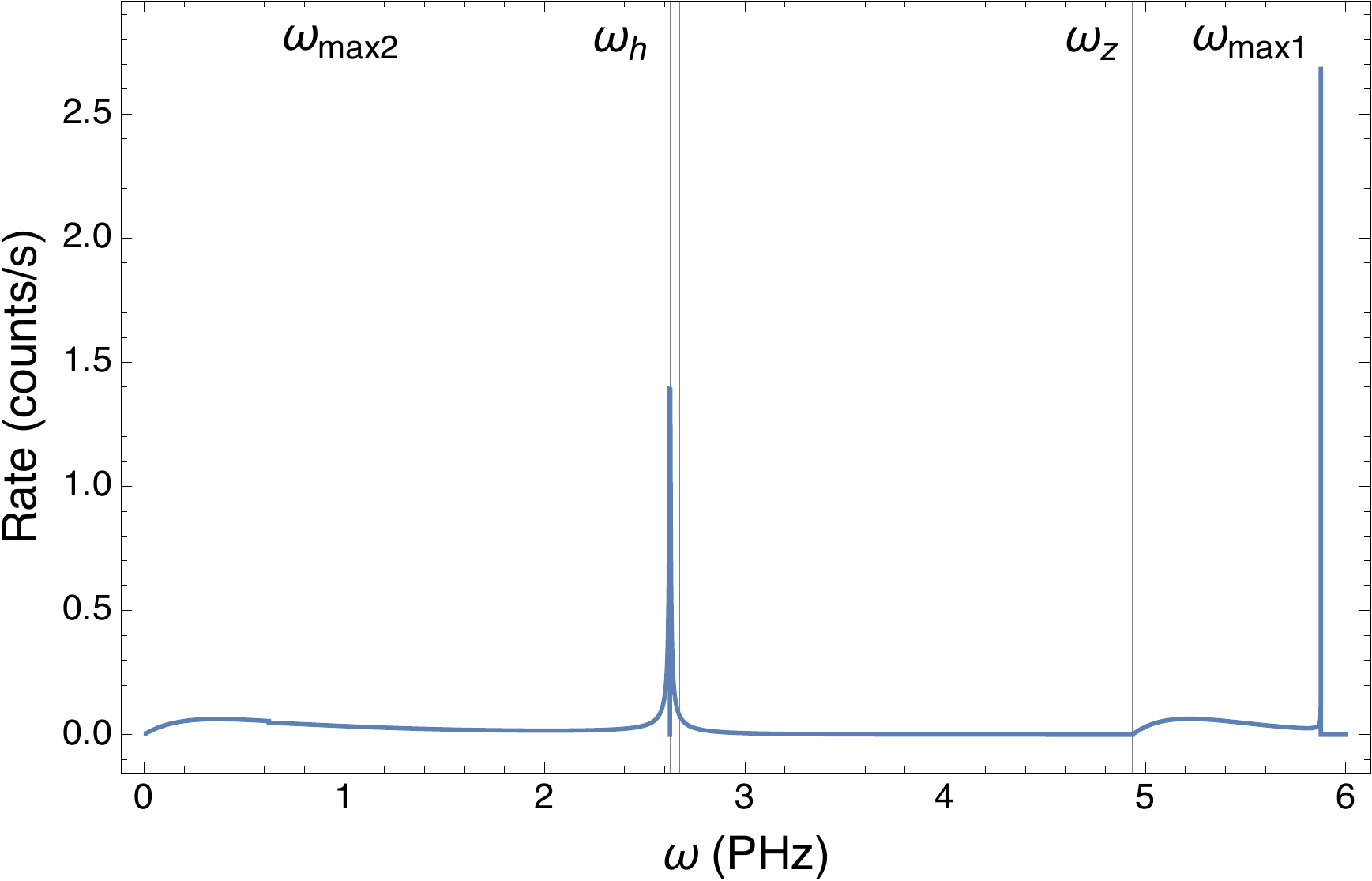}
	\caption{(Color online). Full Hawking spectrum for the dispersion. The peaks are in $\omega_h$ and $\omega_{\text{max1}}$. The vertical gray lines mark the shown frequencies, the lines on the sides of $\omega_h$ are $\omega_{\text{min3}}$ and $\omega_{\text{min3'}}$. Also, to the right of $\omega_\text{max2}$ is $\omega_\text{min2}$ and to the left of $\omega_\text{max1}$ is $\omega_\text{min1}$, both indistinguishable from this scale. These lines limit the region for creation of Hawking radiation.}
	\label{fighsfull}
\end{figure}

\begin{figure}
	\centering
	\includegraphics[width=85mm]{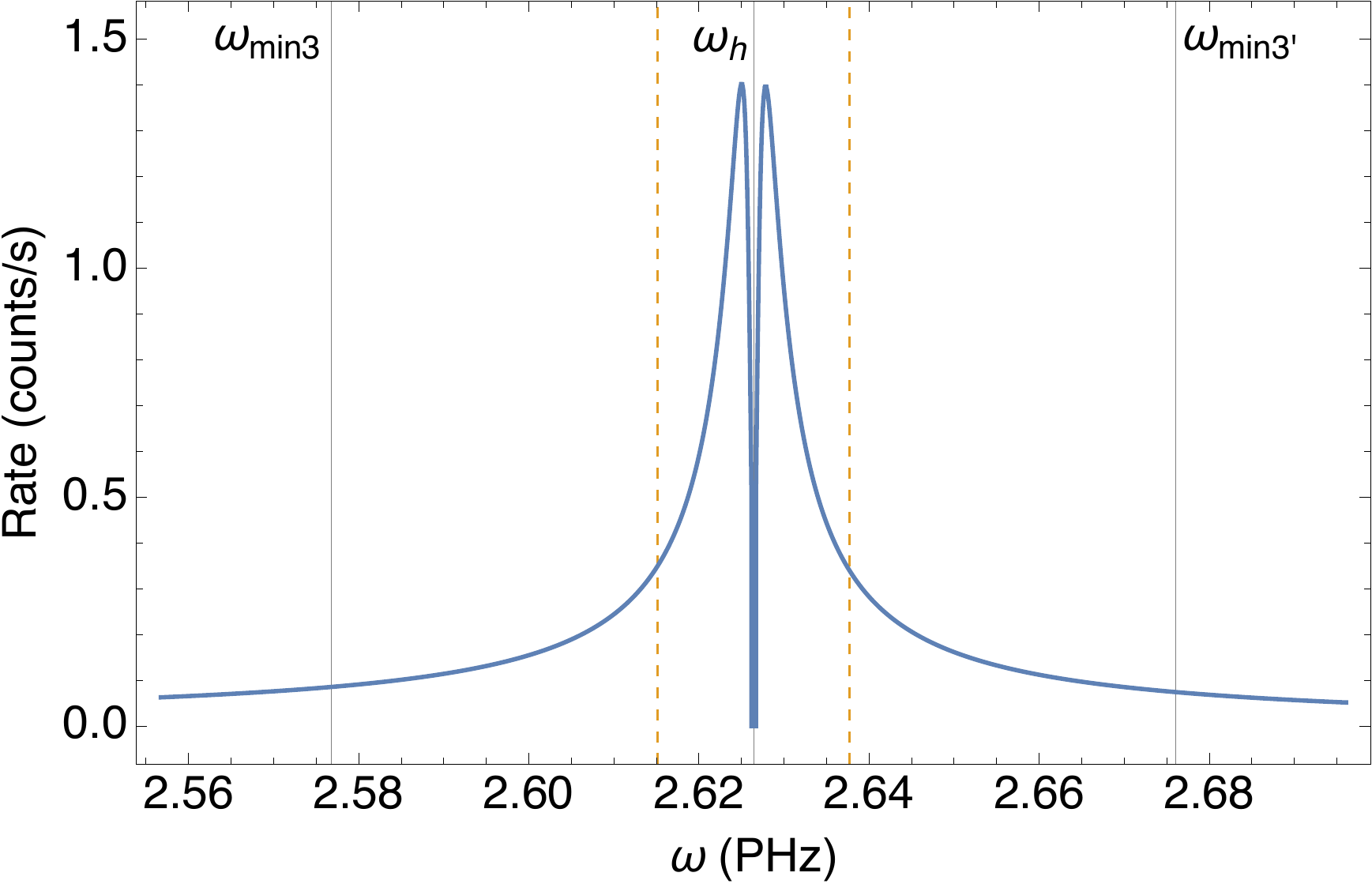}
	\includegraphics[width=85mm]{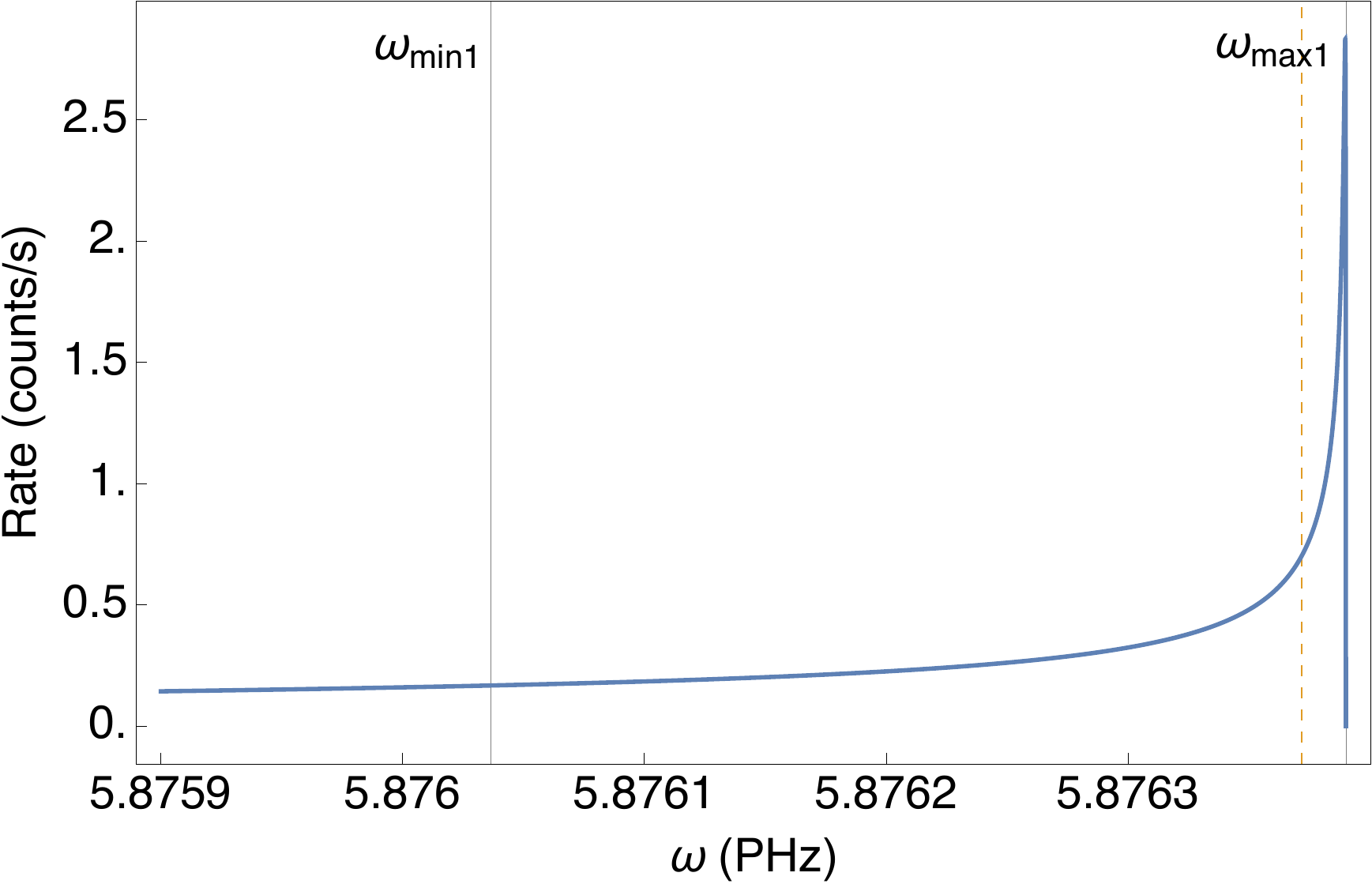}
	\caption{(Color online). Close up of the Hawking spectrum around the horizon in $\omega_h$ and around the negative frequency horizon in $\omega_\text{max1}$. In both cases we observe that the particle production dips to zero exactly at the horizon. The dashed orange lines correspond to the same arbitrary value of $\omega'$ and they will be correlated.}
	\label{fighsclose}
\end{figure}

The full Hawking spectrum for the dispersion from Sec. \ref{method} and a soliton pulse given by $\chi_0=10^{-3}$ and FWHM time of 2\,fs is shown in Fig. \ref{fighsfull}. This spectrum has two main features. First, even though in a first approximation the only regions where Hawking radiation could be produced are the ones shown at the end of Sec. \ref{study} and in Fig. \ref{figwpwclose}, this spectrum contains Hawking radiation outside those regions. This result is somehow expected, as the Hawking production should be continuous: the emission rate is much lower outside the expected region, but not zero. On the other hand, the Hawking spectrum is higher around the horizon in $\omega_h$ and it presents a clear dip exactly there, i.e., the production of Hawking radiation is exactly zero at the horizon. At first, this seems contradictory but it also follows from the theory in Sec. \ref{stasfluid}. As can be seen from Eq. \eqref{approx}, the relative derivative is the essential quantity for the production rate of Hawking radiation in an analog system for the optical case, while the change of speed $u$ is given by the Kerr effect $\delta n$ in Eq. \eqref{n2a}. For the soliton, the derivative starts from zero away from the pulse and increases slowly up to a maximum point and then decreases rapidly to zero only at the horizon, which explains the dip in the spectrum in Fig. \ref{fighsfull}. We show a close up to the spectrum around the horizon in the top part of Fig. \ref{fighsclose}.

The Hawking radiation in the UV region presents similar features going to zero in $\omega_h$, as seen in Fig. \ref{fighsfull}. In this case we do not have the dip, because mode 1 only has one branch. These effect can be seen in the close up to mode 1 in the bottom part of Fig. \ref{fighsclose}.

Of course this Hawking radiation will have the usual properties expected from its production of quantum origin, i.e., it will be quantized and the Hawking spectrum will eventually be build up from an accumulation of photons (like in the classic double-slit experiment with single photons). Also, we must remember that $\omega'$ is the conserved quantity, therefore all particles created by the Hawking mechanism at a given time will have the same value in $\omega'$. This presents as an entanglement between all particles with the same $\omega'$ created at the same time. Thus, one simple way of verifying if the emitted radiation is really Hawking radiation is to measure its correlations. Particles from mode 1 ranging from $\omega_{\text{min1}}$ to $\omega_{\text{max1}}$ will be correlated with those from mode 3 from $\omega_{\text{min3}}$ to $\omega_h$ and those from mode $3'$ from $\omega_{\text{min3'}}$ to $\omega_h$. For example, the orange lines in both sides of Fig. \ref{fighsclose} show the corresponding $\omega$ in the modes 3, $3'$, and 1 for the same $\omega'$ that lies inside the region of creation of Hawking radiation. In theory, the same happens for mode 2, but as we saw, it is very small because of its propagation in the opposite direction. A strong test of the quantumness of the radiation is to check if photons at those frequencies are correlated and entangled, which is something routinely performed in quantum-optics laboratories.

\subsection{Analysis of numerical errors} \label{errors}
As we pointed out earlier, whenever we use a numerical method to solve a differential equation the question of stability becomes important. We checked the numerical method in two ways: we examined the stability of the results while changing the grid spacing and the norm conservation of scattered waves.

In the first case, the main parameter that governs the stability of the algorithm is the number of points in the $\omega'$ region (or conversely, the grid spacing combined with the limits of the grid). We tried several number of grid points until the point where adding or subtracting some of them will make no difference on the results.

\begin{figure}
	\centering
	\includegraphics[width=85mm]{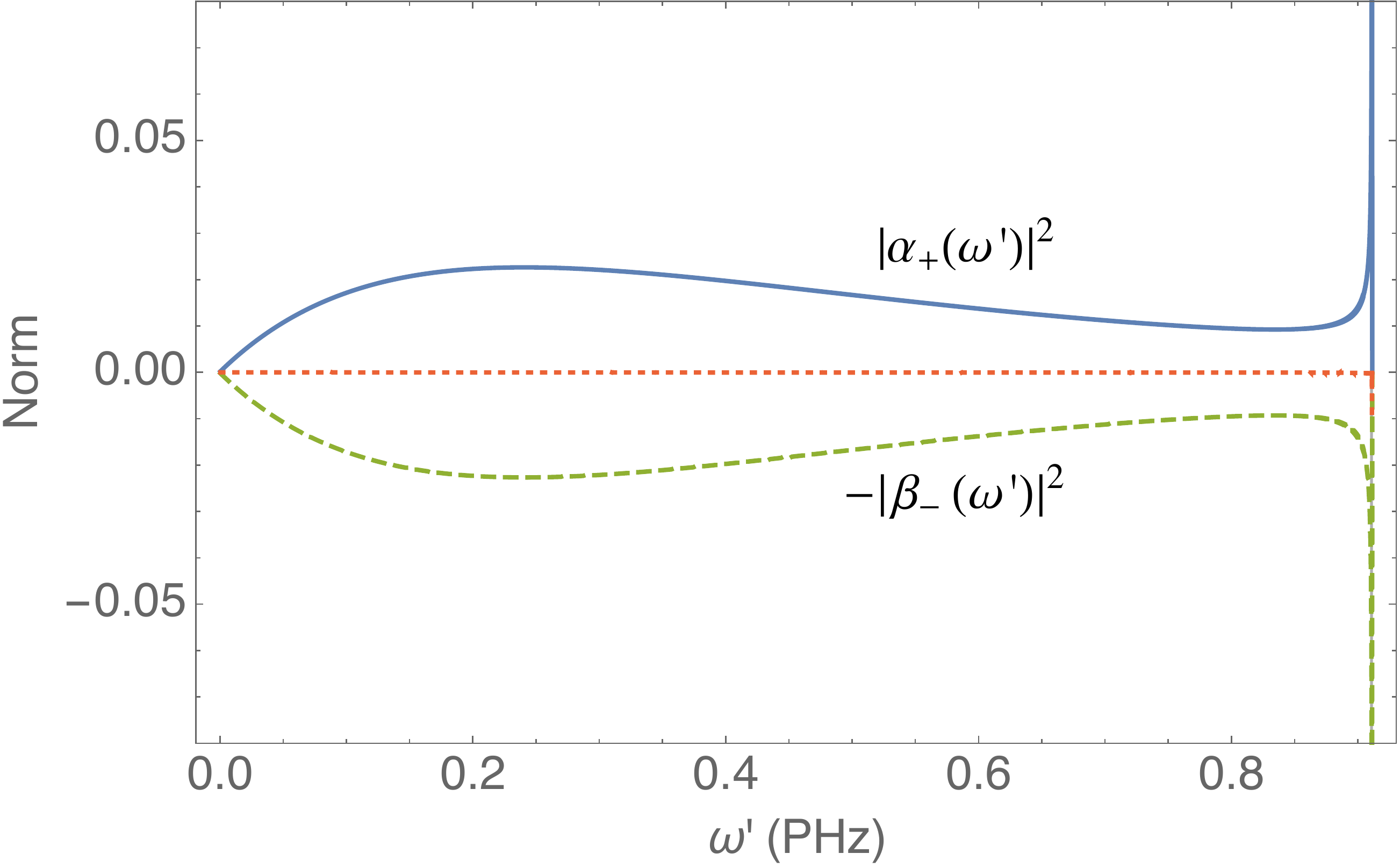}
	\caption{(Color online). Norm conservation. The positive-norm (solid blue) is the sum of the norm of modes 2, 3, and $3'$, the negative norm (dashed green) is the norm of mode 1. We also show the sum of both of them and we check the norm conservation (dotted red). The norm is normalized according to the maximum (close to the horizon).}
	\label{fignorm}
\end{figure}

For the second point, as we mentioned in Sec. \ref{scattering} and in Eq. \eqref{normcon}, the norm should be conserved during the scattering process. Furthermore, in cases when there are negative-norm waves, this process is more like a charge conservation. So, from the physical point of view, the conservation of norm gives us a very strong test of our method, specially in the region close to the horizon. We obtain the sum of positive-norm modes $\alpha_+(\omega')$ (norms 2, 3, and $3'$ from Fig. \ref{figwpwfull}) from the negative-norm mode (mode 1) $\beta_-(\omega')$:
\begin{subequations}
	\begin{align}
	|\alpha_+(\omega')|^2 &= |\alpha(\omega';\omega_2)|^2+|\alpha(\omega';\omega_3)|^2+|\alpha(\omega';\omega_{3'})|^2,\\
	|\beta_-(\omega')|^2&=|\beta(\omega';\omega_1)|^2.
	\end{align}
\end{subequations}
The notation in these equations is slightly different from that in Eq. \eqref{normcon}, but remember that, according to Table \ref{table}, in the optical analog the role of $k$ and $\omega$ is played by $-\omega$ and $\omega'$. In Fig. \ref{fignorm} we show the results of this analysis. The norm is plotted in terms of $\omega'$, from zero to $\omega'_h$. We also show the sum of all the norms and see that it stays very close to zero and only increases near the horizon (right-hand side of the plot) although it is still relatively small. All the norms are normalized according to the maximum absolute value of the positive norm (it is almost the same as the negative maximum absolute value). We can also see that the emission of Hawking radiation outside the horizon is possible although much smaller than inside.

Another important figure of merit is the relative difference between the positive and negative norms [the relative error $E_r(\omega')$], because the spectrum has most of the radiation inside the horizon region. Given the normalization in the previous plot we have:
\begin{equation}
E_r(\omega')=\frac{|\alpha_+(\omega')|^2-|\beta_-(\omega')|^2}{|\beta_-(\omega')|^2}.
\end{equation}

\begin{figure}
	\centering
	\includegraphics[width=85mm]{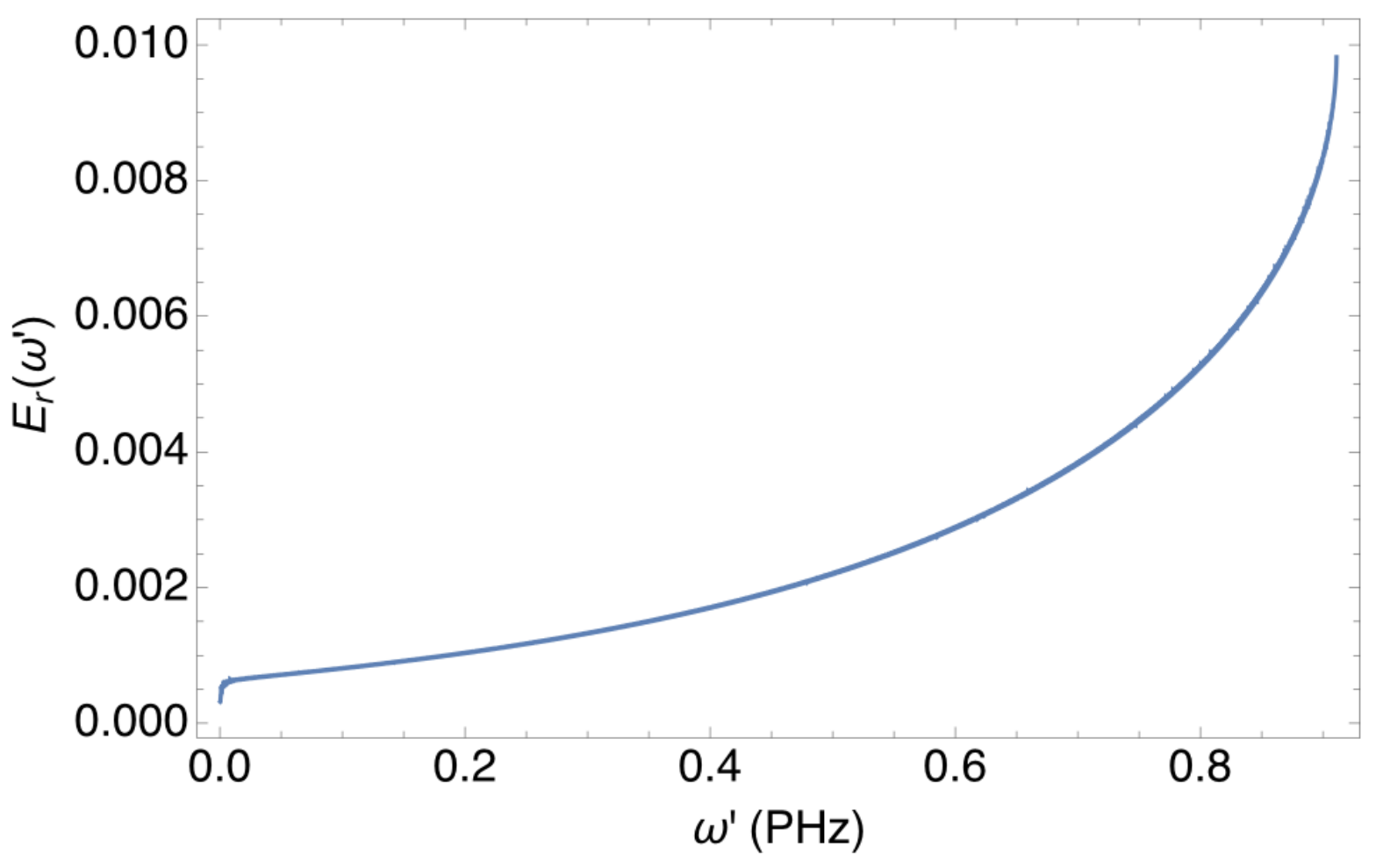}
	\caption{(Color online). Relative difference between the positive- and negative-norm modes (relative error) in terms of $\omega'$. As expected, the maximum value is at the horizon and it is less than 1\%.}
	\label{figerror}
\end{figure}

In Fig. \ref{figerror} we plot $E_r(\omega')$ and we find a continuous line that increases outside the horizon and it stays close to the previous values in the horizon. The maximum value is less than 1\% and it is in the point of maximum creation of particles.

\section{Conclusions}\label{conclusions}
In this work we have presented a complete method to calculate the Hawking spectrum for a system in quantum optics. To do this, we must know the dispersion relation of the material in which the effect is expected: bulk material, waveguide, or optical fiber. Also, the shape of the pulses traveling inside it, usually a soliton.

We have shown that the emission of Hawking radiation is dominated by the group-velocity horizon. Most of the radiation of the Hawking spectrum is emitted in the spectral region of the horizon. Even though the horizon is fuzzy in space, it is clearly defined in Fourier space and its study is easier.

We presented here the calculation for a dispersion given by a theoretical approximation from an optical waveguide. Nevertheless, we can apply this method to any dispersion relation; for example for those materials listed in Ref. \cite{refindex}. Also, we must remark that in this case we consider a soliton that keeps its shape during its travel through the material. When this is the case, such pulse creates both black-hole and white-hole horizons and our method does not distinguish between their radiations. We study the scattering process of the pulse as a whole, therefore this spectrum is characteristic of the black-hole/white-hole pair. In particular, the central dip is a characteristic fulfilled for each system, so the creation of particles will always go to zero at the horizon.

We have seen that Hawking radiation is not bounded by frequencies reaching the horizon, i.e., the scattering process creates photons even outside that part of the spectrum, although it greatly amplifies the emission rate. Simulations without the horizon or with a long pulse (weak horizon) also create particles although with a rate $10^{-8}$ smaller. This means that the process is still valid, but without the horizon there is no amplification.

This work gives a theoretical framework to experimental groups working to achieve the quantum-optics analog of the black-hole horizon and to produce Hawking radiation in those systems. We have obtained a system that works in the optical regime and has well separated signals for the Hawking partners in the laboratory frame, $\omega_h$ and $\omega_\text{max1}$. For this dispersion, they are in $\sim$717 nm and $\sim$321 nm, respectively, i.e., in the IR and UV and close to the visible range; it is not hard to find efficient detectors in those ranges.

We are confident in the results presented here, given the complexity of the algorithm, its stability, and relative errors.  We trust that the features shown can be expected in a real experiment and can serve as a theoretical tool to design the experiment itself.

It would also be interesting to study this system under the effects of a non-symmetric pulse, which could be obtained by the effect of nonlinearities that cause self-steepening and increase the derivative on one side of the pulse and decrease it on the other. In this case, we would expect an increase in the Hawking spectrum on one side of the double peak around $\omega_h$ and a decrease on the other. This system has some advantages, as the higher derivative will boost the production of photons and will narrow its range of frequencies; therefore, the production of photons would be higher and it would also be easier to measure their entanglement. On the other hand, the pulse dynamics that causes self-steepening usually breaks the pulse very fast (in distances of the order of millimeters); therefore, the interaction cannot be as long as in the case of the soliton. An interesting question is what are the optical conditions to have the maximal production of Hawking radiation that is experimentally measurable.

\begin{acknowledgments}
D.B. would like to thank Scott Robertson and Jonathan Drori for valuable discussions. D.B. also thanks the people at Weizmann Institute of Science for their hospitality during part of the period in which this work was done. This work was supported by the European Research Council, the Israeli Science Foundation, Conacyt (Mexico) Project 152574, and a research grant from Mr. and Mrs. Louis Rosenmayer and from Mr. and Mrs. James Nathan.
\end{acknowledgments}

\appendix*
\section{}\label{integral}
In this Appendix we will detail the integral method used to calculate the scattering matrix $\mathcal{S}$, which was then applied to find the Hawking spectrum. This Appendix follows the method by Robertson and Leonhardt \cite{RL14}.

\subsection{Wave equation}
Let us take the general metric given by Eq. \eqref{metric}, which describes the motion in a fluid where $c=c(-i\partial_x )$ is the speed of waves, which is not constant for dispersive media, and $u(x)$ is here the velocity of the fluid. Then, the scalar wave-equation is
\begin{equation}
[\partial_t +\partial_x u(x)][\partial_t + u(x)\partial_x]\phi - c^2(-i\partial_x) \partial_x^2\phi =0.
\end{equation}
The dispersion makes us lose the exact analogy with the fluid but it will help us model our system more realistically. In particular, it will also test how strong is the hypothesis of Hawking radiation in the dispersion problem. First, let us solve the time-dependent part by imposing a harmonic dependence $\phi (x,t) = \exp (-i\omega t) \phi_\omega (x)$ to get:
\begin{equation}
[-i \omega +\partial_x u(x)][-i\omega + u(x)\partial_x]\phi_\omega - c^2(-i\partial_x) \partial_x^2\phi_\omega =0,
\label{waveq}
\end{equation}
Now we perform the Fourier transform of this equation. As usual, we denote $\widetilde\phi_\omega$ the Fourier transform of $\phi_\omega$ (from now on we drop the subscript $\omega$ in the notation):
\begin{equation}
\widetilde\phi(k)=\int_{-\infty}^{+\infty} \exp(-ikx) \phi (x) \dd x.
\end{equation}
If we transform Eq. \eqref{waveq} to the Fourier space, the terms with $u(x)$ give rise to convolutions and the existence of spatial derivatives to more complicated terms. The resulting equation is:
\begin{equation}
g(k)\widetilde\phi(k)+\int_{-\infty}^{+\infty} K(k,k') \widetilde\phi(k') \dd k'=0,
\label{integraleq}
\end{equation}
where
\begin{subequations}
	\begin{align}
	g(k)&=c^2(k)k^2-\omega,\\
	K(k,k')&= \frac{1}{2\pi}[\omega(k+k')\widetilde{u}(k-k')-kk' \widetilde{u^2}(k-k')],
	\end{align}
\end{subequations}
where $\widetilde u$ and $\widetilde{u^2}$ are Fourier transforms.

\subsection{Half Fourier transforms}\label{halfft}
On the other hand, let us split the function $\widetilde\phi(k)$ into two with respect to the zero of the spatial coordinate $x$, which gives rise to what is called half Fourier transforms:
\begin{subequations}
	\begin{align}
	\widetilde\phi^L(k) &=\int_{-\infty}^0\phi(x)\exp(-ikx) \dd x,\\
	\widetilde\phi^R(k) &=\int_0^{\infty}\phi(x)\exp(-ikx) \dd x,
	\end{align}
\end{subequations}
where the $L,R$ superscripts refer to left-hand and right-hand side integrals and $\phi$ should be asymptotically bounded. We can also split the kernel $K(k,k')$ with the condition that it is smooth in $x=0$ as $K(k,k')=K^L(k,k')+K^R(k,k')$. This has the advantage that some of the terms will vanish and the equations become easier to handle. Then Eq. \eqref{integraleq} gives:
\begin{align}
&g(k)[\widetilde\phi^L(k)+\widetilde\phi^R(k)]+\int_{-\infty}^{+\infty} K_L(k,k') \widetilde\phi^L(k') \dd k'\nonumber\\
&+\int_{-\infty}^{+\infty} K_R(k,k') \widetilde\phi^R(k') \dd k'=0.
\end{align}
Furthermore, we can decompose $\widetilde\phi^L(k)$ and $\widetilde\phi^R(k)$ into its singular and regular parts. The singular part depends on the half Fourier transforms of plane waves $\exp(i k_\omega x)$, given by:
\begin{subequations}
	\begin{align}
	\frac{1}{2\pi}\int_{-\infty}^0 \exp(ik_\omega x) \exp(-ikx) &=\frac{1}{2}\delta (k-k_\omega)-\frac{1}{2\pi i} \mathcal{P} \frac{1}{k-k_\omega},\\
	\frac{1}{2\pi}\int_0^{\infty} \exp(ik_\omega x) \exp(-ikx) &=\frac{1}{2}\delta (k-k_\omega)+\frac{1}{2\pi i} \mathcal{P} \frac{1}{k-k_\omega},
	\end{align}
\end{subequations}
where $\mathcal{P}$ is the principal part. Therefore, if we define $\alpha^L$ and $\alpha^R$ as the regular parts of $\widetilde\phi^L $ and $\widetilde\phi^R$, respectively; they indicate the short-range behavior in the scattering region. Then we have
\begin{subequations}
	\begin{align}
	\widetilde\phi^L(k) &=\alpha^L (k)+\sum_{j=1}^{N_L}\mathcal{A}_j^L\left[ \frac{1}{2}\delta (k-k^L_\omega)-\frac{1}{2\pi i} \mathcal{P} \frac{1}{k-k^L_\omega} \right],\\
	\widetilde\phi^R(k) &=\alpha^R (k)+\sum_{j=1}^{N_R}\mathcal{A}_j^R\left[ \frac{1}{2}\delta (k-k^R_\omega)+\frac{1}{2\pi i} \mathcal{P} \frac{1}{k-k^R_\omega} \right],
	\end{align}
\end{subequations}
where $\mathcal{A}_j^L$ and $\mathcal{A}_j^R$ are the corresponding amplitudes. Hence, using the analyticity properties of the half kernels, we obtain an integral equation for $\alpha = \alpha^L + \alpha^R$:
\begin{align}
&g(k)\alpha(k)+\int_{-\infty}^{+\infty} K_L(k,k') \alpha(k') \dd k'\nonumber\\
&+\sum_{j=1}^N \mathcal{A}_j^L\left[-\frac{1}{2\pi i}\frac{g(k)}{k-k_j}+K_L(k,k_j)\right]\nonumber \\
&+\sum_{j=1}^N \mathcal{A}_j^R\left[-\frac{1}{2\pi i}\frac{g(k)}{k-k_j}+K_R(k,k_j)\right]=0,
\end{align}
where we keep the terms with $L$ and $R$ separated, as it will be convenient next. Following Robertson and Leonhardt \cite{RL14}, we have the following set of regularity conditions for $\alpha(k)$
\begin{align}
&\int_{-\infty}^{+\infty} K(k_i,k') \alpha(k') \dd k'\nonumber\\
&+\sum_{j=1}^N \mathcal{A}_j^L\left[-\frac{1}{2\pi i}g'(k_i) \delta_{ij}+K_L(k_i,k_j)\right]\nonumber \\
&+\sum_{j=1}^N \mathcal{A}_j^R\left[-\frac{1}{2\pi i}g'(k_i)\delta_{ij}+K_R(k_i,k_j)\right]=0.
\label{regcon}
\end{align}
We can regularize this equation by removing the zeros of $g(x)$ and then by dividing by it, to obtain:
\begin{align}
&\alpha(k)+\int_{-\infty}^{+\infty} \bar{K}_L(k,k') \alpha(k') \dd k'\nonumber\\
&+\sum_{j=1}^N \mathcal{A}_j^L\bar{K}_L(k,k_j)
+\sum_{j=1}^N \mathcal{A}_j^R\bar{K}_R(k,k_j)=0,
\label{alphak}
\end{align}
where the overbars indicate the subtraction of the regularity conditions, i.e., they are defined by
\begin{equation}
\bar{K}(k)=\frac{K(k)}{g(k)}-\sum_{j=1}^N \frac{K(k_j)}{(k-k_j)g'(k_j)}.
\end{equation}
There is an invertible kernel from the left-hand side of Eq. \eqref{alphak} such that we can clear $\alpha(k)$ as
\begin{equation}
\alpha(k)=-\int_{-\infty}^{+\infty}V(k,k')\sum_{j=1}^N
\left[ \mathcal{A}_j^L\bar{K}_L(k,k_j) + \mathcal{A}_j^R\bar{K}_R(k,k_j)\right]\dd k'.
\end{equation}
Then, we substitute $\alpha$ into the regularity conditions of Eq. \eqref{regcon} to obtain the following set of equations:
\begin{equation}
\mathcal{M}_L\vec{\mathcal{A}}^L+\mathcal{M}_R\vec{\mathcal{A}}^R=0,
\label{mama}
\end{equation}
where we have defined:
\begin{subequations}
	\begin{align}
	[\mathcal{M}_L]_{ij} =&-\frac{1}{2\pi i}g'(k_i)\delta_{ij}+ K_L(k_i,k_j)\nonumber\\
	&-\iint_{-\infty}^\infty K(k_i,k)V(k,k')\bar{K}_L(k',k_j)\dd k' \dd k,\\
	[\mathcal{M}_R]_{ij} =&\frac{1}{2\pi i}g'(k_i)\delta_{ij}+ K_R(k_i,k_j)\nonumber\\
	&-\iint_{-\infty}^\infty K(k_i,k)V(k,k')\bar{K}_R(k',k_j)\dd k' \dd k.
	\end{align}
\end{subequations}
These equations describe the relation between amplitudes of the asymptotic plane waves, as we wanted, but they are written as left- and right-hand sides. In order for this result to be applicable, we must re-write them in the in- and out-base. In order to do that, we define a variable $s_j=\pm 1$, which is positive or negative according if $k_j$ is outgoing to the left or the right, respectively. Moreover, we define projection operators $\mathcal{Q}^\pm$ such that they keep only the corresponding values. Therefore, the amplitudes in the in and out bases are:
\begin{subequations}
	\begin{align}
	\vec{\mathcal{A}}^\text{out} & =\mathcal{Q}^- \vec{\mathcal{A}}^L + \mathcal{Q}^+ \vec{\mathcal{A}}^R,\\
	\vec{\mathcal{A}}^\text{in} & =\mathcal{Q}^+ \vec{\mathcal{A}}^L + \mathcal{Q}^- \vec{\mathcal{A}}^R.
	\end{align}
\end{subequations}
Let us define $\mathcal{M}=-(\mathcal{M}_R)^{-1}\mathcal{M}_L$ so that Eq. \eqref{mama} gives
\begin{equation}
\vec{\mathcal{A}}^R =\mathcal{M}\vec{\mathcal{A}}^L.
\end{equation}
Then, using the properties of the projection operators $\mathcal{Q}^\pm$ and after some algebra we get
\begin{align}
(\mathbb{1}-\mathcal{Q}^- \mathcal{M}^{-1}\mathcal{Q}^+ &- \mathcal{Q}^+\mathcal{M}\mathcal{Q}^-)\vec{\mathcal{A}}^\text{out} =\nonumber\\
&(\mathcal{Q}^- \mathcal{M}^{-1}\mathcal{Q}^- + \mathcal{Q}^+\mathcal{M}\mathcal{Q}^+)\vec{\mathcal{A}}^\text{in},
\end{align}
which means that the scattering matrix $\mathcal{S}$ in Eq. \eqref{matrixS} is given by
\begin{align}
\mathcal{S}=&(\mathbb{1}-\mathcal{Q}^- \mathcal{M}^{-1}\mathcal{Q}^+ - \mathcal{Q}^+\mathcal{M}\mathcal{Q}^-)^{-1}\nonumber\\
&\times (\mathcal{Q}^- \mathcal{M}^{-1}\mathcal{Q}^- + \mathcal{Q}^+\mathcal{M}\mathcal{Q}^+).
\end{align}

\subsection{Notes on implementation}
In order to use this method numerically, first, we must obtain the discretized version of the quantities defined there, which can be done straightforwardly. We consider a static pulse with soliton shape and we use the Fourier and half Fourier transforms from Sec. \ref{secpulse}. In addition, we use the dispersion relation from Sec. \ref{study}. Also, the discretization is performed in $\omega'$ and we can see in Fig. \ref{figwpwfull} that there are four different modes that must be solved for each $\omega'$, their labels are also shown. These modes fulfill:
\begin{equation}
s = (-1, +1, -1, -1),
\end{equation}
where $s=+1$ for copropagating and $s=-1$ for counterpropagating waves. Also, 
\begin{equation}
\sigma = (+1,+1, -1, +1),
\end{equation}
where $\sigma=+1$ for waves out-going to the right and $\sigma=-1$ for the ones out-going to the left. Then, for each one of them we obtain the matrix elements $\mathcal{S}^{-1}$ through the integral method. As usual, the grid spacing should be chosen with care, as the code breaks down after some large enough value. Fortunately, the code is fast enough to be run with high precision.

\bibliography{Refs}

\end{document}